\documentclass[fleqn,usenatbib]{mnras}

\usepackage{newtxtext,newtxmath}
\usepackage{orcidlink}

\usepackage[T1]{fontenc}
\usepackage{multirow}
\usepackage{xspace}
\usepackage{balance}
\usepackage[dvipsnames]{xcolor}

\hypersetup{
    colorlinks = true,
    citecolor = {MidnightBlue},
    linkcolor = {BrickRed},
    urlcolor = {BrickRed}
}

\DeclareRobustCommand{\VAN}[3]{#2}
\let\VANthebibliography\thebibliography
\def\thebibliography{\DeclareRobustCommand{\VAN}[3]{##3}\VANthebibliography}
\def\HI{\textsc{Hi}\xspace}
\newcommand\edit[1]{#1}

\usepackage{graphicx}	

\title[Number counts and clustering of \HI galaxies]{Predicted number counts and clustering of \HI galaxies from future \\ radio surveys}

\author[A. Nasirudin, P. Bull  \& I. Ye]{
Ainulnabilah Nasirudin$^{1}$,\thanks{E-mail: ainulnabilah.nasirudin@manchester.ac.uk} \orcidlink{0000-0003-2213-4547}
Philip Bull\,\orcidlink{0000-0001-5668-3101}$^{1,2}$,
and Isabelle Ye\,\orcidlink{0009-0007-1958-3364}$^{1}$
\\
$^{1}$Jodrell Bank Centre for Astrophysics, University of Manchester, Manchester, M13 9PL, United Kingdom\\
$^{2}$Department of Physics and Astronomy, University of Western Cape, Cape Town 7535, South Africa\\
}

\date{Accepted XXX. Received YYY; in original form ZZZ}

\pubyear{\the\year{}}

\begin{document}
\label{firstpage}
\pagerange{\pageref{firstpage}--\pageref{lastpage}}
\maketitle

\begin{abstract}
The 21cm emission line from neutral hydrogen (\HI) contained within galaxies provides a way to make accurate spectroscopic redshift determinations {\edit at radio frequencies. Large arrays such as SKA-MID will have the sensitivity required to catalogue millions of \HI galaxies.} The expected number counts and clustering properties of the galaxies are still quite poorly understood {\edit however, leading to uncertainties in the predicted performance of cosmological surveys}. We use three different simulated galaxy catalogues to predict the properties of the \HI galaxy distribution, along with estimates of the error on these predictions due to modelling uncertainty. {\edit The simulations are} S$^3$-SAX (semi-analytic models based on the Millennium dark matter-only simulation); GAEA (an updated semi-analytic model partially calibrated on hydrodynamical simulations); and IllustrisTNG (a hydrodynamical simulation). We present predictions for galaxy number counts as a function of sensitivity and redshift, and use these to forecast the cosmological performance of a proposed SKA-MID cosmological survey. {\edit The simulated predictions differ by a factor of $\sim 2$ in the number counts and around 10\% in the bias at $z \sim 0.1$, yielding around 20\% differences in forecast cosmological uncertainties. These differences become significantly larger at $z \gtrsim 0.4$}. {\edit We also model a `medium-deep' SKA \HI galaxy survey at $z \lesssim 0.07$ by fitting a halo occupation distribution (HOD) model to angular correlation functions measured from multiple $20^\circ \times 20^\circ$ sub-fields of IllustrisTNG. This provides clustering predictions and uncertainty estimates that can be used in future forecasts}.
\end{abstract}

\begin{keywords}
galaxies: haloes -- cosmology: cosmological parameters -- cosmology: large-scale structure of Universe
\end{keywords}



\section{Introduction}



Galaxies form within the cosmic web, and as a result they (imperfectly) trace the spatial distribution of the dominant dark matter component. The relationship between observable galaxies and the underlying dark matter is typically described using halo models, in which statistical relationships are defined between particular types of galaxy and generic dark matter haloes characterised by their mass \citep{Jing_1998, 10.1046/j.1365-8711.2000.03779.x, COORAY20021, 2002ApJ...575..587B, 10.1093/mnras/stw845, 2017MNRAS.471...12B, 2023OJAp....6E..39A}. These relationships have free parameters that are then calibrated using observations or simulations.

On large scales, halo models can be used to model and predict the linear bias, which quantifies the relative strength of clustering of the galaxies with respect to the dark matter field. On small scales, the clustering models become more complex, and must statistically describe the spatial distribution of galaxies within individual haloes. These two regimes -- large scales, and sub-halo scales -- are described by the 2-halo and 1-halo terms respectively. A common approach is to use a `Halo Occupation Distribution' (HOD) to model these terms \citep{2002ApJ...575..587B}. This specifies the mean number of galaxies -- possibly divided into central and satellite populations \citep{2005ApJ...633..791Z} -- as a function of the mass of the host halo. Alternative approaches, such as Sub-Halo Abundance Matching (SHAM) \citep{Kravtsov_2004, 2006ApJ...647..201C, 10.1093/mnras/stw845}, are also available.

The HOD parameters can be inferred observationally by fitting to two-point functions of the galaxy number count distribution, i.e. quantities that describe the covariance of the galaxy density field as a function of separation or distance scale. A variety of two-point functions can be formed depending on the available information about galaxy positions \citep[for a summary, see][]{Chisari_2019}; for instance, if the 3D {\edit (redshift-space)} positions of galaxies are accurately known through spectroscopic redshift measurements, the correlation function $\xi(\vec{r})$ (in configuration space) or power spectrum $P(\vec{k})$ (in Fourier space) can be {\edit inferred}. For galaxies with less precise radial positions (e.g. photometric redshift estimates), it is more common to measure the angular correlation function $w(\vec\theta)$ (in angular separation) or the angular power spectrum $C_\ell$. Cross-spectra between pairs of redshift bins may also be formed. All of these quantities can be predicted through appropriate projections of the galaxy power spectrum, which can in turn be calculated by applying the HOD model to the dark matter power spectrum \citep{Chisari_2019}.

At optical and near-infrared wavelengths, large galaxy catalogues have been constructed and used to make high signal-to-noise measurements of two-point functions, permitting HOD parameters to be inferred for various galaxy populations at a range of redshifts \citep[e.g.][]{2020MNRAS.499.5486A, 2024MNRAS.530..947Y, 2025ApJ...983..183S}. The HOD parameters that are recovered depend on the properties of the galaxy samples, such as their depth, colour selections, and/or photometric redshift binning, but can be used to gain insight into the physical clustering properties of the galaxies and the masses of the dark matter haloes that they are embedded in.

The clustering properties of radio galaxy populations are typically less well constrained, {\edit although measurements have been made for a number of different galaxy samples, ranging from continuum galaxies and AGN hosts \citep[e.g.][]{2021MNRAS.502..876A, 2024MNRAS.527.6540H, 2026MNRAS.547ag468H}, to \HI galaxies \citep[e.g.][]{2013ApJ...776...43P, 2017ApJ...846...61G, 2019MNRAS.486.5124O}.} Part of the problem is the difficulty of obtaining precise redshifts for large samples of radio galaxies. Those detected through continuum (i.e. synchrotron and free-free) emission lack well-defined spectral features that could be used to estimate redshifts and so must be cross-matched with other redshift tracers \citep[e.g.][]{Jackson:1998pr, 2017arXiv170408278H}. Only a handful of emission/absorption lines are readily observable at GHz frequencies and below, and these tend to be relatively faint. The most notable radio emission line for the purposes of this work is the 21cm line from neutral hydrogen (\HI). Neutral hydrogen is ubiquitous in galaxies, tending to exist in dense self-shielded regions where it is not easily ionised by background UV photons. The line itself tends to be rather faint however, and so the only large samples of \HI galaxies to date are at very low redshifts -- the ALFALFA catalogue of 31,500 sources extends to $z \lesssim 0.06$ for example \citep{2018ApJ...861...49H}.

We are now entering an era of rapid growth in the sensitivity of radio arrays, and hence the size and depth of the galaxy surveys that they can perform \citep{2015A&ARv..24....1G, 2015aska.confE.167S}. For the purposes of survey planning and theoretical modelling, it is useful to be able to predict the expected number density of galaxies as a function of redshift and flux density, and the corresponding clustering properties of the sample \citep[e.g.][]{2015aska.confE..21S, 2017MNRAS.469.2323P, 2018ApJ...866..135V}.

These quantities are quite uncertain for \HI galaxies however, and so the clearest predictions for fainter and higher-redshift samples tend to come from simulations. These are subject to substantial model uncertainties {\edit however. To date, cosmological \HI galaxy survey predictions for SKA have mostly relied on single simulations, some of which used cosmological parameters that are now outdated \citep[e.g.][]{yahya10.1093/mnras/stv695}. There can also be substantial differences in predicted number counts and galaxy bias predictions between simulations, driven by differences in mass resolution, volume, and neutral gas modelling for example. This is most problematic at higher redshifts, $z \gtrsim 0.4$, where significantly different survey strategies and sample properties would be implied depending on the choice of simulation.}

In this paper, we make updated predictions for the \HI galaxy number counts that will be achievable with current and near-future radio arrays, including ASKAP, MeerKAT, and SKA-MID \citep{Maddox2021, Rhee2023, KazemiMoridani2025}. These arrays are capable of measuring substantial numbers of \HI galaxies out to $z \gtrsim 1$ depending on the survey strategy, and can also catalogue large numbers of \HI galaxies over thousands of square degrees at much lower redshifts, $z \lesssim 0.3$ \citep{2020PASA...37....7S}. As such, the \HI galaxy population predictions must reach flux densities of order $\mu$Jy, and redshifts beyond unity.

Our approach is to compare the \HI galaxy populations in three simulated datasets -- the original S$^3$-SAX semi-analytic catalogue based on the Millennium dark matter-only simulation \citep{Obreschkow_2009b, yahya10.1093/mnras/stv695, Bull:2015lja}; an updated semi-analytic simulation called GAEA, which is partly calibrated using hydrodynamical simulations \citep{2017MNRAS.465.2236Z, mayor_2026}; and the IllustrisTNG hydrodynamical simulation \citep{tng2018MNRAS.475..624N, 2019MNRAS.484.5499S}. While hardly exhaustive -- {\edit other suitable simulations are available, such as SIMBA \citep{2019MNRAS.486.2827D} } -- this gives a handle on the level of theoretical uncertainty that exists when simulating this kind of galaxy in a cosmological context. To help put these numbers in context, we use the S$^3$-SAX and GAEA predictions to forecast the uncertainties on cosmological distance measurements (the expansion rate and angular diameter distance) as a function of redshift for the planned SKA-MID configurations.

We then take a high-resolution $z \approx 0$ snapshot of IllustrisTNG and use it to predict the clustering properties of the \HI galaxies at the low redshifts where wide-area surveys will be possible. Rather than working directly on the simulated matter distribution, we compute the angular correlation function from a mock catalogue of \HI galaxies with a simple flux density threshold selection {\edit to provide more observationally-relevant predictions of the clustering observables}. We then use a Bayesian parameter inference approach to fit the HOD model parameters to the simulated observations \citep{ye-acf10.1093/mnras/staf1651}. By dividing the snapshot into nine sub-regions, each of size $20^\circ \times 20^\circ$ and redshift range $0.01 \le z \le 0.067$ {\edit (similar to, but slightly shallower than a proposed SKA-MID `medium-deep' survey), we are also able to study the variance in predictions between fields, and verify the robustness of empirical covariance matrix estimates.}

This paper is arranged as follows. Section \ref{sec:survey_specs} shows the calculation of the expected sensitivity of the SKA-MID instrument with different array configurations. In Section \ref{sec:hi_sims}, we provide details about the \HI simulations we use and the post-processing steps involved. Next, in Section~\ref{sec:cosmo_fisher}, we discuss the cosmological constraints from Fisher matrix forecasts run with the \HI observables. We then present the effect of sample variance on the angular correlation function and HOD model in Section \ref{sec:acf_hod} before concluding in Section \ref{sec:conclusion}.

\begin{table*}
\begin{tabular}{|c|c|c|c|c|c|}
\hline
Configuration        & Band {[}MHz{]} & $A_{\rm eff} / T_{\rm sys}$ {[}m$^2$/K{]} & Beam FoV {[}deg$^2${]} & $t_p$ {[}hr{]} & Flux dens. rms {[}$\mu$Jy{]} \\ \hline
\multirow{2}{*}{AA*} & 350 - 1050     & 733.2                                     & 1.78                   & 3.55           & 235.2                  \\ \cline{2-6} 
                     & 950 - 1760     & 1201.4                                    & 0.47                   & 0.95           & 277.8                  \\ \hline
\multirow{2}{*}{AA4} & 350 - 1050     & 1029                                      & 1.78                   & 3.55           & 167.6                  \\ \cline{2-6} 
                     & 950 - 1760     & 1733                                      & 0.47                   & 0.95           & 192.6                  \\ \hline
\end{tabular}
\caption{Survey specifications for a total observation time of 10,000 hours and survey area of 5,000 deg$^2$. The values for $A_{\rm eff} / T_{\rm sys}$, primary beam field of view, time per pointing $t_p$, and flux density rms (per beam) are given at the central frequency of the band.}
\label{tbl:survey_specs}
\end{table*}

\section{Telescope and Survey Specification}
\label{sec:survey_specs}
In this section, we calculate the expected sensitivity for detecting the 21cm line from individual galaxies based on the SKA-MID receiver and array specifications. See \cite{yahya10.1093/mnras/stv695} for further details.

The root mean square (rms) flux density measured by a radio interferometer in the limit of a large number of dishes is given by
\begin{equation}
    S_{\rm rms} \simeq \frac{2 k_{\rm B} T_{\rm sys}}{A_{\rm eff} \sqrt{2 \,\delta \nu \, t_p}} \, \textrm{Jy/beam}.
\end{equation}
Here, $k_{\rm B}$ is the Boltzmann constant, $T_{\rm sys}$ is the system temperature, $A_{\rm eff}$ is the total effective collecting area, $\delta \nu$ is the width of the frequency channel, and $t_p$ is the observation time per pointing. {\edit This is the flux density rms in each frequency channel. Scaling by values representative of modern arrays}, the equation can be re-written as
\begin{equation}
\label{eqn:fluxrms}
    S_{\rm rms} = 260 \, \mu\textrm{Jy}/\textrm{bm} \left( \frac{T_{\rm sys}}{20 \textrm{\, K}}\right)
    \left( \frac{25,000 \textrm{\,m}^2}{A_{\rm eff}}\right) \left( \frac{10\, \textrm{kHz}}{\delta \nu}\right)^{1/2} \left( \frac{1 \textrm{hr}}{t_p}\right)^{1/2}.
\end{equation}
If the total useful observing time for a survey, $t_{\rm tot}$, is divided uniformly across a {\edit survey area $\Omega_{\,{\rm survey}}$,} $t_p$ can be calculated as
\begin{equation}
    t_p \approx \frac{t_{\rm tot}}{\Omega_{\,{\rm survey}} }  \left[\frac{\pi}{8} \left( \frac{1.3 \lambda }{D} \right)^2 \right].
\end{equation}
Here, $\lambda$ is the wavelength, and $D_{\rm dish}$ is the dish diameter, where the value in the square bracket is the area per pointing equal to the area of the primary beam as a function of wavelength. With SKA-MID, for a 5,000~deg$^2$ survey over 10,000~hours \citep{2020PASA...37....7S}, $t_p$ corresponds to 3.55 and 0.95 hours at the central frequency of Band 1 (350 -- 1050 MHz) and Band 2 (950 -- 1760 MHz) respectively.

We use values of $A_{\rm eff} / T_{\rm sys}$ from the anticipated SKA1 science performance memo\footnote{\url{https://www.skao.int/sites/default/files/documents/SKAO-TEL-0000818-V2_SKA1_Science_Performance.pdf}} to calculate the flux density rms using Equation~\ref{eqn:fluxrms} for SKA1-MID AA* and AA4 configurations\footnote{The AA* configuration consists of 144 dishes with a maximum baseline of 108.0 km, while the AA4 configuration consists of of all 197 dishes with a maximum baseline of 159.6 km.} for one hour per pointing. A summary of the resulting survey specifications is given in Table~\ref{tbl:survey_specs}.

\section{\HI Galaxy Simulations}
\label{sec:hi_sims}

In this section, we outline the conversion steps used on the IllustrisTNG simulation output to get to the observable quantities, along with calculations of the \HI galaxy number density and bias from all three simulations.

 The following selection cuts have been applied to all simulations, following \cite{yahya10.1093/mnras/stv695}:
\begin{itemize}
    \item Observed \HI line width $W_\theta > 2 \delta V$;
    \item Total \HI flux density $S > N_{\rm cut}\, \sigma_S$.
\end{itemize}
{\edit The flux density $S$ should be interpreted as a mean flux density per channel over the width of the line, i.e. $S = \mathcal{F} / W_\theta$, where $\mathcal{F}$ is the velocity-integrated flux of the line and $W_\theta$ is the observed line width, uncorrected for inclination. The error on the mean flux density is $\sigma_S = S_{\rm rms} / \sqrt{W_\theta / \delta V}$, assuming that the noise rms in each channel adds in quadrature. In what follows, we set $N_{\rm cut}=5$, and take the frequency channel resolution to be 10 kHz, corresponding to a velocity width of $\delta V= 2.1 (1+z)$ km s$^{-1}$. This is a relatively optimistic $5\sigma$ threshold for source detection, but a more conservative cut on line width that excludes (mostly face-on) galaxies that do not have the characteristic broad double-horned line profile \citep{Obreschkow_2009b}. }

\subsection{\HI Galaxy Observables from IllustrisTNG}
\label{sec:hi_conversion}

IllustrisTNG\footnote{\url{https://www.tng-project.org/}} is a suite of state-of-the-art cosmological magnetohydrodynamical simulations of galaxy formation and evolution \citep{tng2018MNRAS.475..624N, tng2018MNRAS.475..648P, tng2018MNRAS.475..676S, tng2018MNRAS.477.1206N, tng2018MNRAS.480.5113M, tng2019ComAC...6....2N}. There are three physical box sizes that probe different astrophysical processes: 35, 75, and 205 Mpc/$h$, commonly referred to as TNG50, TNG100, and TNG300 respectively. For this work, we use the supplementary data post-processed from the original catalogue, which provide the \HI and H$_2$ content of the galaxies at snapshot redshifts $z = [0, 0.5, 1.0, 1.5, 2.0]$ \citep{diemer2019tng}.

\begin{figure*}
    \centering
    \includegraphics[width=0.98\linewidth]{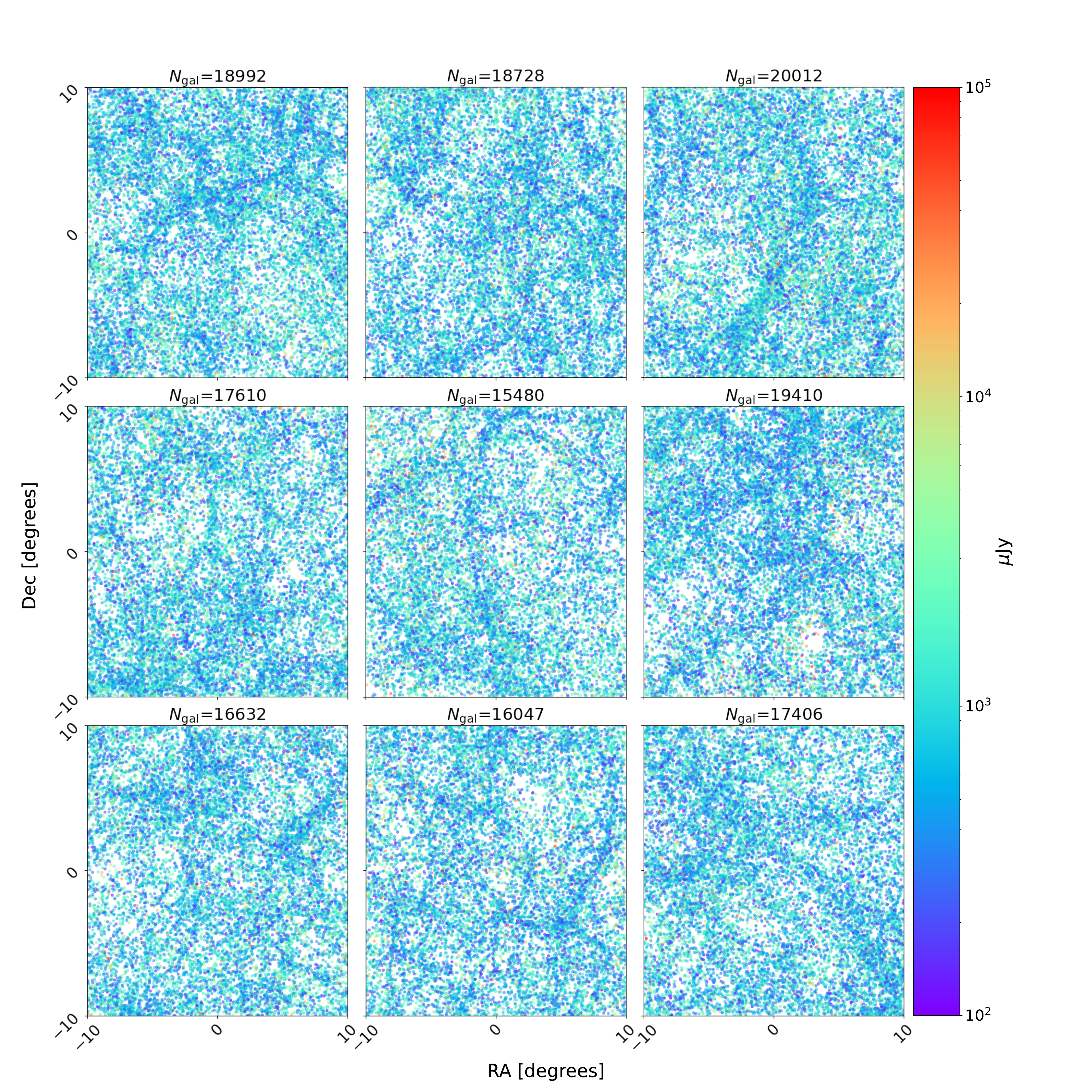}
    \caption{Distribution of \HI galaxies in {\edit $20^\circ \times 20^\circ$} chunks of the TNG300 simulation box, for galaxies between $0.01 \leq z \leq 0.067$. The galaxies are colour-coded according to their observed flux density for an observer at a fixed reference position in the simulation box. The total number of galaxies with flux density above $100\ \mu$Jy is stated at the top of each panel.}
    \label{fig:hi_galaxies} 
\end{figure*}

\begin{figure*}
    \centering
    \includegraphics[width=0.49\linewidth]{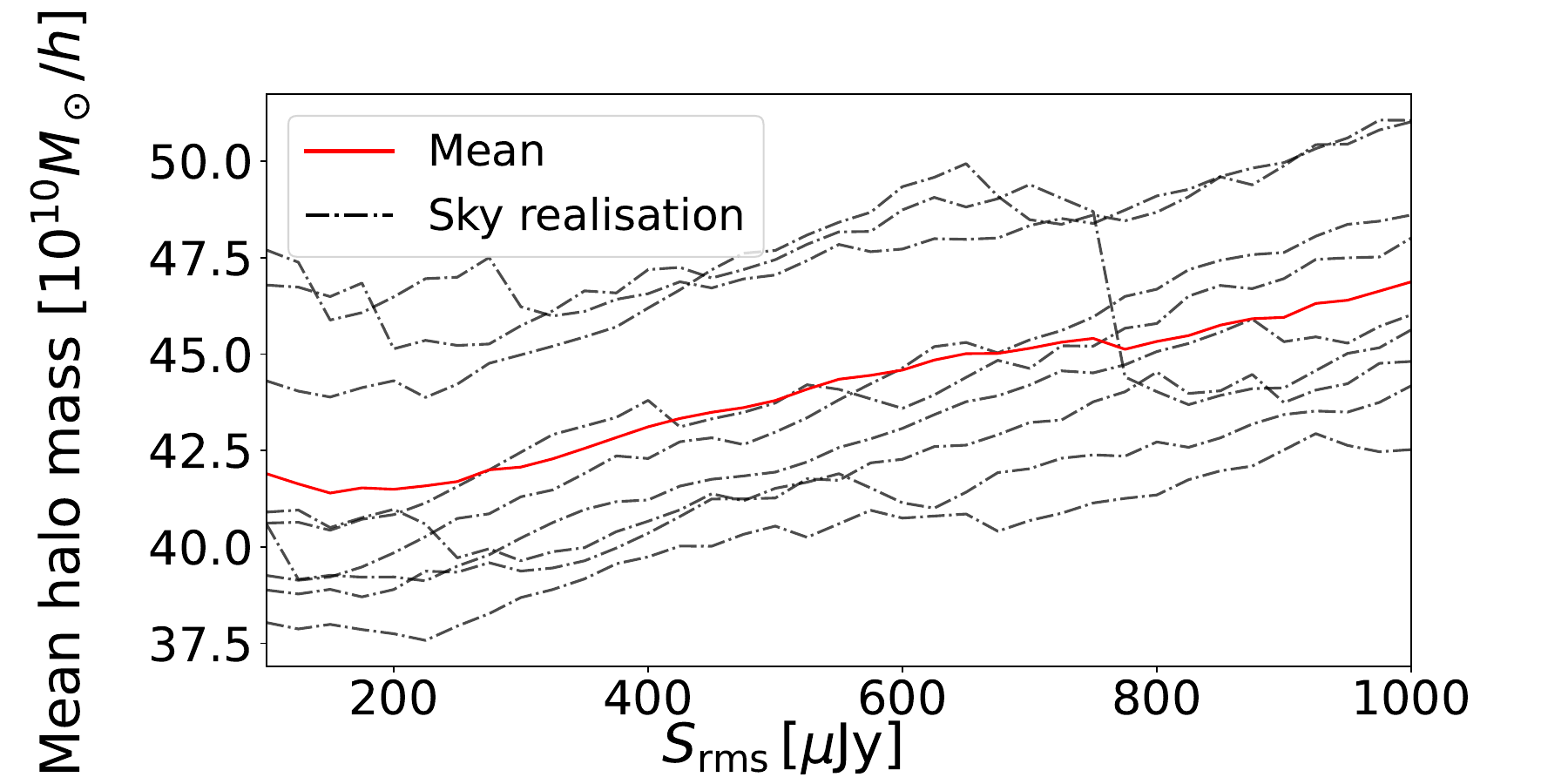}
    \includegraphics[width=0.49\linewidth]{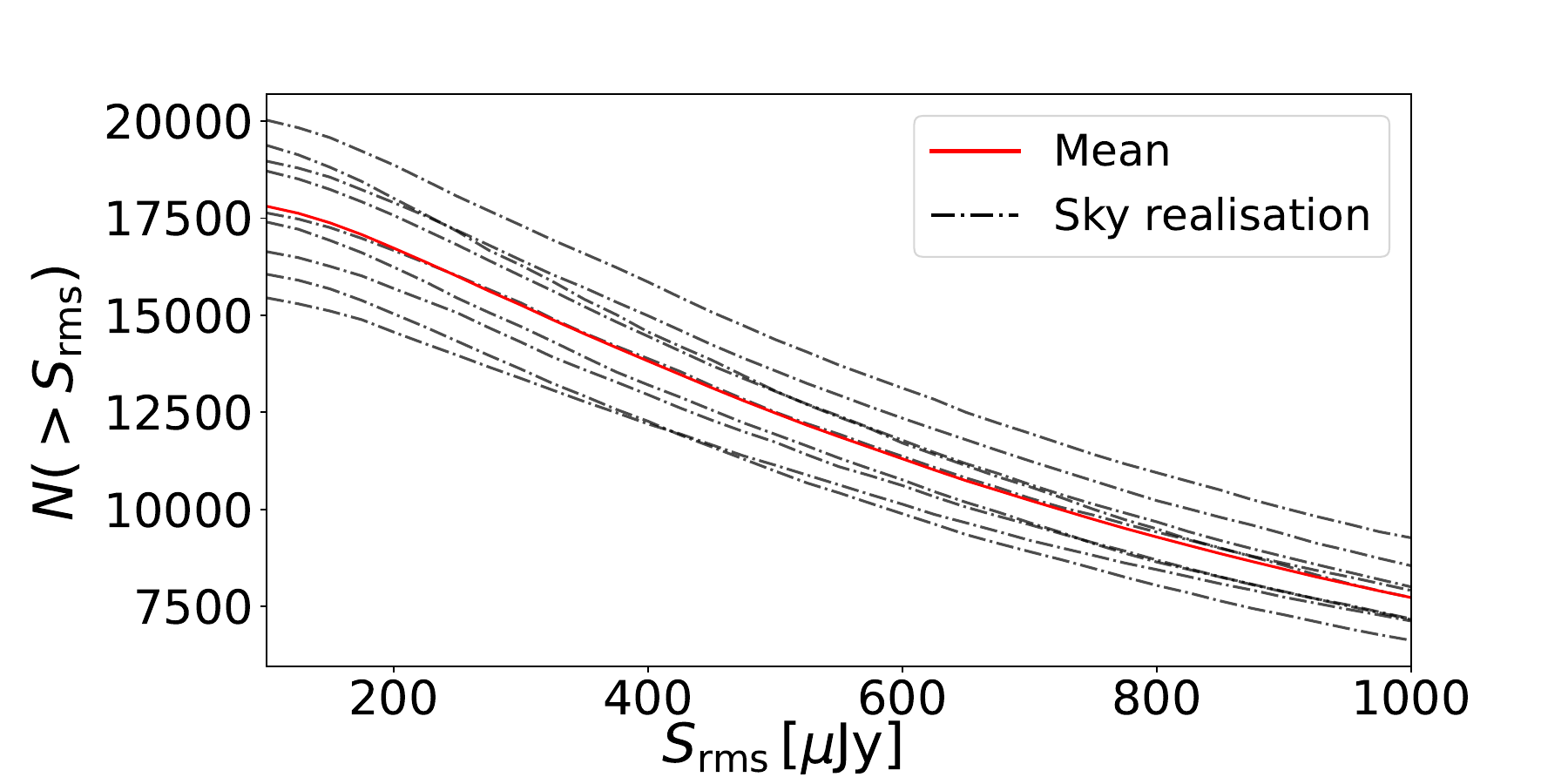}
    \caption{The mean halo mass (left) and cumulative number of galaxies {\edit with mean flux in each channel, $S$, above a flux density threshold $N_{\rm cut} \sigma_S$ (right) as a function of flux density rms} cut for the nine different fields. The dash-dotted lines are the values for each field, while the solid line represents the mean across all nine fields.}
    \label{fig:meanhmass_ngal}  
\end{figure*}

\begin{figure}
    \centering
    \includegraphics[width=\linewidth]{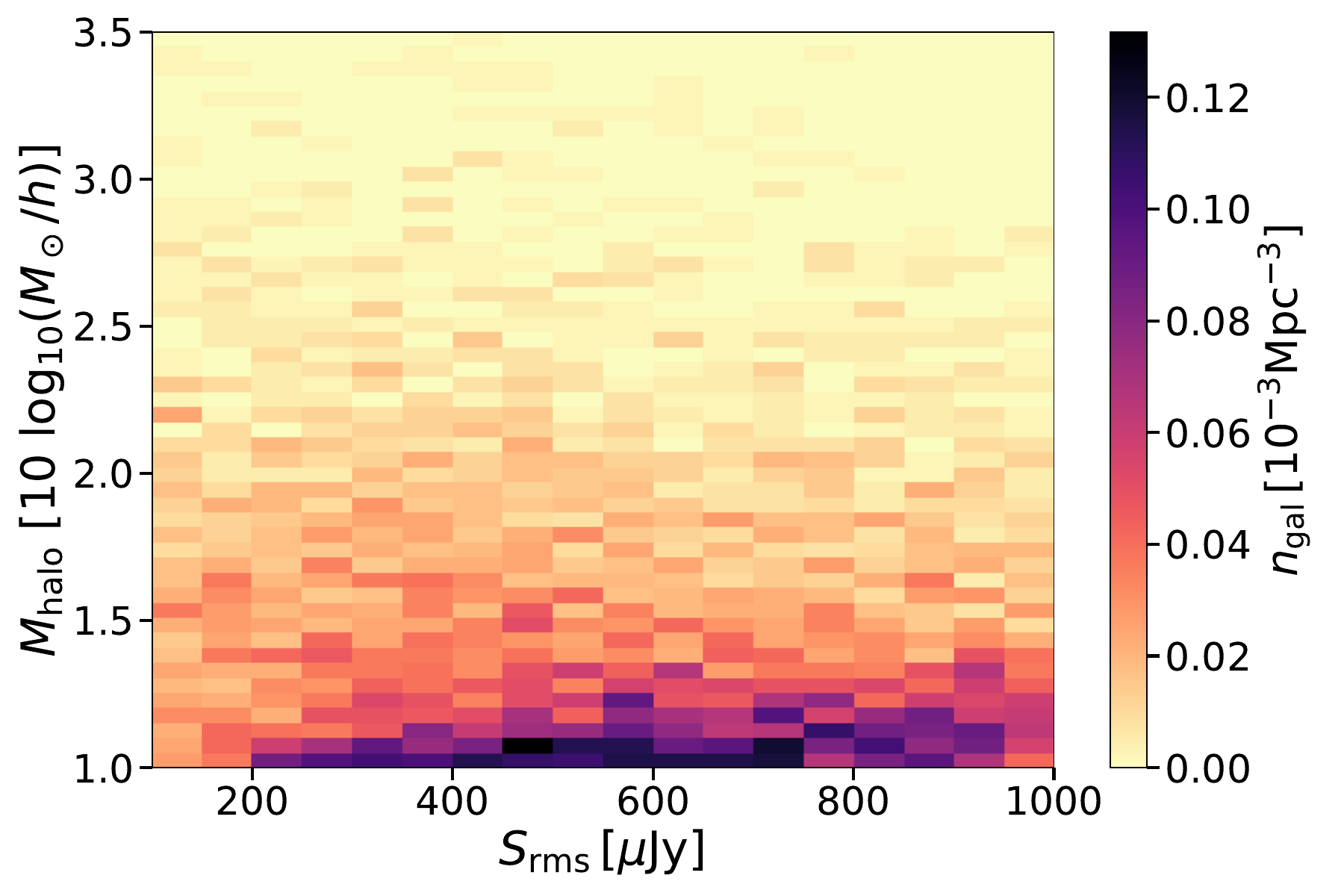}
    \caption{The number of galaxies as a function of halo mass and flux density threshold for one of the nine fields. At $S_{\rm rms}=800 \, \mu$Jy, there is notable gap in the number of high-mass haloes.}
    \label{fig:mhalo_flux} 
\end{figure}

We convert the atomic \HI mass, $M_{\rm \HI}$, to the intrinsic \HI line width, $W_{\rm e}$, following the relationship
\begin{equation}
    \frac{W_{\rm e}}{420 \textrm{\, km s}^{-1}} = \left(\frac{ M_{\rm \HI}}{10^{10} \textrm{\, M}_\odot}\right)^{0.3}
    \label{eqn:mass_width}
\end{equation}
from \cite{duffy2012askap}. Because details on the projected inclination of these galaxies are not included in the TNG catalogue, we use the same method as \cite{duffy2012askap} whereby a value of inclination angle $\theta$ is assigned to each \HI galaxy following $\cos\theta \sim {\rm U}(0,1)$.
The line width of an inclined galaxy, $W_\theta$, can then be calculated as
\begin{equation}
    (W_{\rm e} \sin\theta)^2 = W_\theta^2 + V_{\rm o}^2 - 2 W_\theta V_{\rm o} \left[1 - e^{-\left(\frac{W_\theta}{V_{\rm c}}\right)^2}\right] - 2 V_{\rm o}^2 e^{-\left(\frac{W_\theta}{V_{\rm c}}\right)^2},
    \label{eqn:tullyfouque}
\end{equation}
where $V_{\rm c}=120$ km s$^{-1}$ marks a transition between the Gaussian vs `boxy' line profiles seen in small vs large galaxies respectively; and $V_{\rm o} \approx 20$ km s$^{-1}$ is a contribution from random motions within the disc \citep{1985ApJS...58...67T, duffy2012askap}.

If we assume that the \HI galaxies are optically thin, the observed-frame, {\edit velocity-integrated flux $\mathcal{F}$} is then related to $M_{\rm \HI}$ by
\begin{equation}
    M_{\rm \HI} \simeq \frac{2.35 \times 10^5\, h^{-2} {\rm M}_\odot}{(1+z)^2} \left( \frac{D_L}{h^{-1} \rm{\,Mpc}} \right)^2 \left(\frac{\mathcal{F}}{\rm{Jy \, km \, s}^{-1}} \right),
\end{equation}
 where $h$ is the dimensionless Hubble constant and $D_L$ is the luminosity distance \citep{Meyer_Robotham_Obreschkow_Westmeier_Duffy_Staveley-Smith_2017}. We can then divide $\mathcal{F}$ by $W_\theta$ to get {\edit the mean flux density across the width of the line, $S$}.

We use two different treatments to calculate the distribution of \HI galaxies from the TNG100 and TNG300 simulations, according to what they will be used for.

\subsubsection{TNG100}
\label{sec:TNG100}
The higher resolution data from TNG100 are used to calculate the \HI galaxy number density, $dN/dz$, and bias, $b(z)$, which are the two important ingredients needed to predict cosmological observables from the detected \HI galaxies. This is because we expect low-mass haloes to host a considerable amount of \HI. Instead of constructing a lightcone, we have chosen to interpolate the values between the snapshots. 
Specifically, we calculate $dN/dz$ following these steps:
\begin{enumerate}
    \item Bin the data according to $z$, $M_{\rm \HI}$, and $W_\theta$;
    \item Divide number count by simulation volume to find $dN/dV(z)$;
    \item Interpolate values of $dN/dV(z)$ between $z=0.01$ and $z=2$;
    \item {\edit Calculate the mean flux density, $S$, from $M_{\rm \HI}$ and $W_\theta$;}
    \item {\edit Apply the two selection cuts on $W_\theta$ and $S$;}
    \item Sum the bins of $dN/dV(z)$ and multiply by $ \frac{dV}{dz d\Omega}$.
\end{enumerate}
The latter quantity is given by
\begin{equation}
    \frac{dV}{dz d\Omega} \, \textrm{[Mpc}^3 \textrm{deg}^{-2} \textrm{ per unit \,} z] = \frac{c r(z)^2}{H(z)} \left(\frac{\pi}{180}\right)^2,
\end{equation}
where $c$ is the speed of light, $r(z)$ is the comoving distance at $z$, and $H(z)$ is the cosmic expansion rate. 

Similarly, we follow the same steps outlined above to calculate the \HI galaxy bias, but with some adjustments:
\begin{enumerate}
    \item Bin the data according to $z$, $M_{\rm \HI}$, $W_\theta$, and halo mass, $M_h$;
    \item Calculate the values of $S$ using the values of $M_{\rm \HI}$ and $W_\theta$;
    \item Apply the two selection cuts;
    \item Calculate the \HI galaxy bias, $b(z, S_{\rm rms})$.
\end{enumerate}
The bias is given by
\begin{equation}
    b(z, S_{\rm rms}) \approx \sum_i b(z, M_h^i) \frac{N(M_h^i)}{N_{\rm tot}},
\end{equation}
where $b(z, M_h^i)$, $N(M_h^i)$, and $N_{\rm tot}$ are the Sheth-Tormen halo bias for mass $M_h^i$ \citep{1999MNRAS.308..119S}, the number of dark matter halo with mass $M_h^i$ passing the selection cuts for a particular $S_{\rm rms}$, and the total number of haloes passing the selection cuts respectively \citep{yahya10.1093/mnras/stv695}.

\subsubsection{TNG300}
\label{sec:TNG300}
We use the data from TNG300 to study the clustering of galaxies (see Section \ref{sec:acf_cov}) because of its larger volume. In this part of our work, we use the data at snapshot 0 and calculate the redshift of a galaxy based on its comoving position from a reference observer. We only include galaxies between $0.01 \leq z \leq 0.067$ with $S\leq 10^5 \mu$Jy. The simulation box is divided into nine different fields, {\edit each with a square field of view of $20^\circ \times 20^\circ$, as shown in Figure~\ref{fig:hi_galaxies}. This area is similar to the `medium deep' SKA \HI galaxy survey tier proposed in \citet{2015aska.confE.167S}, although the latter targets greater depth, with a mean redshift of around 0.2.}

We compare the mean halo mass and cumulative number of galaxies above the flux density threshold as a function of flux density cut from the nine fields in Figure~\ref{fig:meanhmass_ngal}. The dash-dotted lines represent the values in each field while the solid red line represents the mean across all nine fields. The mean halo mass in each field ranges between 3.8 to 4.8 $\times 10^{11} \, M_\odot/h$ at $S_{\rm rms}=100~\mu$Jy. At $S_{\rm rms}=1000~\mu$Jy, this goes up by only about $0.4 \times 10^{11} \, M_\odot/h$, i.e. the mean halo mass changes only slowly with flux density at $z = 0$. The cumulative number of galaxies more than doubles between 1000~$\mu$Jy and 100~$\mu$Jy however.

It is interesting to note that in some fields, the mean halo mass does not increase monotonically as a function of flux density threshold. For example, in one field, the mean halo mass drops drastically around $S_{\rm rms}=800 \mu$Jy. To help understand why, we have included Figure~\ref{fig:mhalo_flux}, showing the number of galaxies as a function of halo mass and flux density threshold for the same field. The sudden drop can be attributed to the lack of rarer high-mass haloes above $S_{\rm rms}=800 \mu$Jy. In contrast, the cumulative number of galaxies decreases monotonically as the flux density threshold increases.

\subsection{Comparing the number density and bias from simulations}
\label{sec:compare_dndz_bz}

\begin{table}
\centering
\begin{tabular}{cccccc}
\hline
$S_{\rm rms}$ [$\mu$Jy] & $c_1$ & $c_2$ & $c_3$ & $c_4$  & $c_5$  \\ \hline
0             & 6.21  & 1.72  & 0.79  & 0.5874 & 9.3577 \\ \hline
1             & 6.55  & 2.02  & 3.81  & 0.4968 & 0.7206 \\ \hline
10            & 6.44  & 1.83  & 7.59  & 0.5928 & 0.8072 \\ \hline
100           & 5.63  & 1.41  & 15.49 & 0.6052 & 1.0859 \\ \hline
200           & 5.00  & 1.04  & 17.52 & --      & --      \\ \hline
\end{tabular}
\caption{Fitting parameter values for $dN/dz$ and $b(z)$ based on the S$^3$-SAX catalogue; taken from \citet{yahya10.1093/mnras/stv695}.}
\label{tbl:params_s3sax}
\end{table}

\begin{table}
\centering
\begin{tabular}{cccccc}
\hline
$S_{\rm rms}$ [$\mu$Jy] & $c_1$ & $c_2$ & $c_3$ & $c_4$  & $c_5$  \\ \hline
0             & 6.370  & 2.117  & 2.609  & 0.738 & 0.297 \\ \hline
1             & 6.339  & 2.102  & 3.091  & 0.738 & 0.300 \\ \hline
10            & 6.538  & 2.208  & 6.432  & 0.723 & 0.442 \\ \hline
100           & 6.849  & 2.336  & 17.895 & 0.710 & 0.923 \\ \hline
200           & 7.170  & 2.492  & 26.385 & 0.716 & 1.086 \\ \hline
\end{tabular}
\caption{Fitting parameter values for $dN/dz$ and $b(z)$ based on the GAEA catalogue\edit{; taken from \citet{Nasirudin01.2026.SKA}.}}
\label{tbl:params_gaea}
\end{table}

\begin{figure*}
    \centering
    \includegraphics[width=1.1\linewidth]{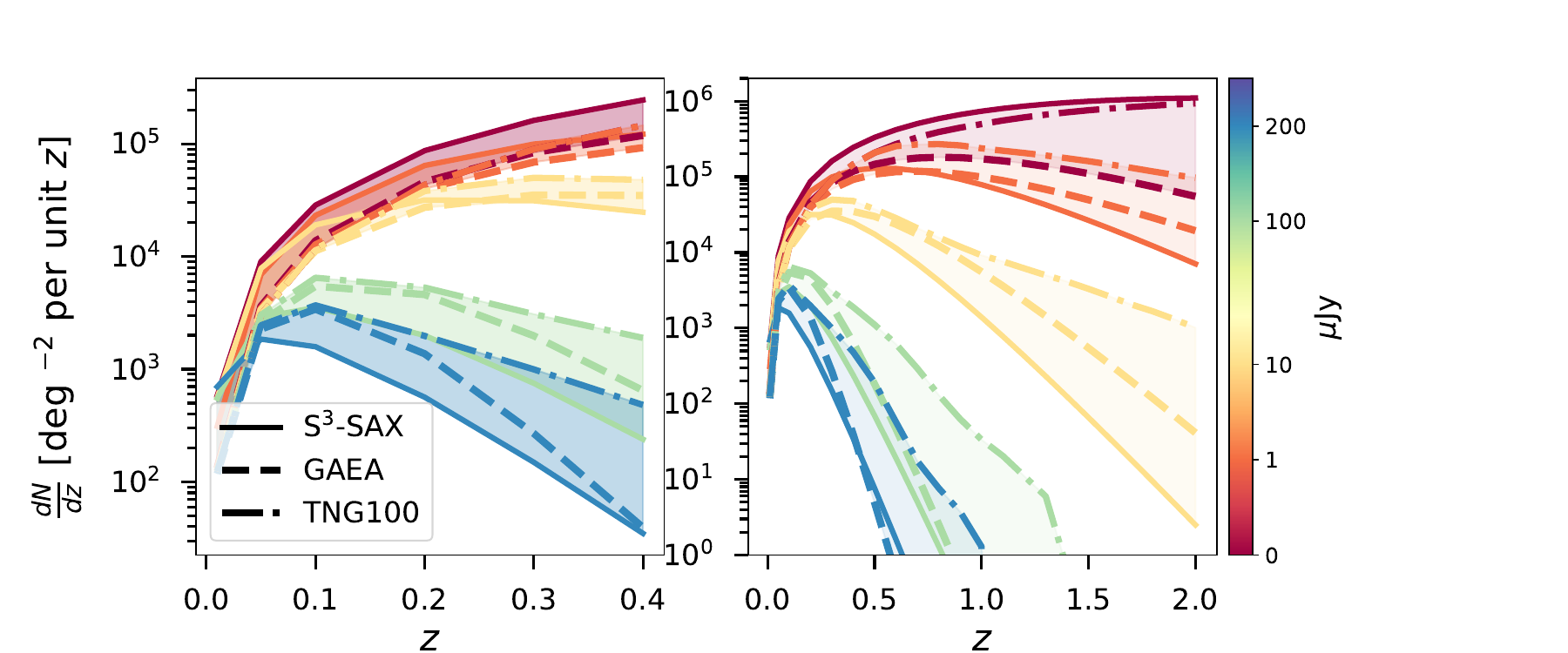}
    \includegraphics[width=1.1\linewidth]{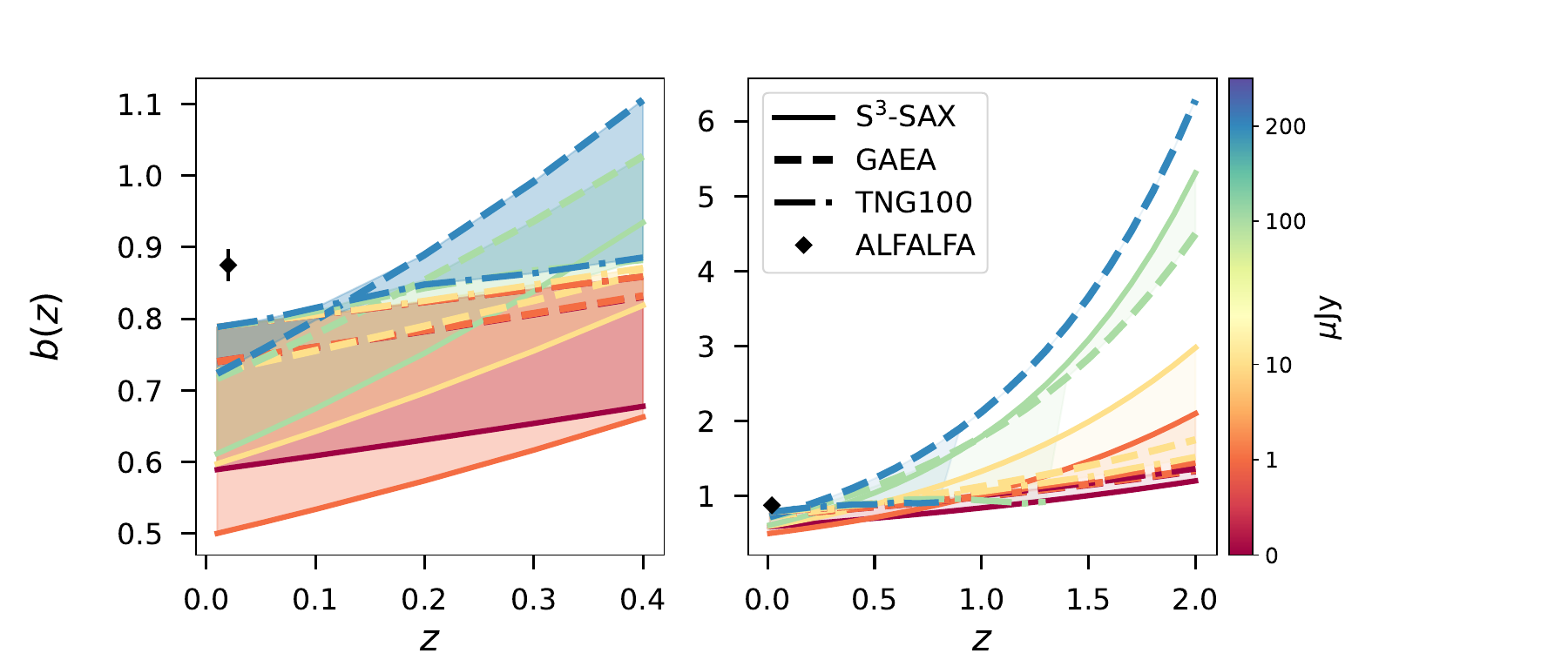}
    \caption{{\it (Top row):} The galaxy number density with different flux density cuts for the S$^3$-SAX (solid), GAEA (dashed), and TNG100 (dash-dot) \HI simulations, {\edit based on the values in Tables~\ref{tbl:params_s3sax} and \ref{tbl:params_gaea}, and the method from Sect.~\ref{sec:TNG100} respectively}. The left panel is a zoom-in of the $z\leq0.4$ range of the right panel. The colours denote the fixed flux rms thresholds according to the colourbar labels. {\it (Bottom row):} The \HI bias with different flux cuts for S$^3$-SAX (solid), GAEA (dash), and TNG100 (dash dot) \HI simulation, along with the observed bias with $S_{\rm rms}=2.4$ mJy from the ALFALFA survey \citep{2019MNRAS.486.5124O}.}
    \label{fig:dndz} \label{fig:bz} 
\end{figure*}

We compare $dN/dz$ and $b(z)$ values as a function of redshift and flux density cut from three different simulations: SKADS Simulated Skies Semi-Analytical eXtragalactic (S$^3$-SAX),\footnote{\url{http://s-cubed.physics.ox.ac.uk/s3_sax}} GAEA, and IllustrisTNG. For the first two simulations, $dN/dz$ and $b(z)$ are calculated from the fitting formulae
\begin{equation} \label{eq:dNdz&bias}
   \frac{dN}{dz} = 10^{c_1} z^{c_2} \exp(-c_3 z); ~~~~~~~
   b(z) = c_4 \exp(c_5 z),   
\end{equation}
where $c_i$ are free parameters that best-fit the simulation data. We provide the values of these parameters in Tables~\ref{tbl:params_s3sax} and \ref{tbl:params_gaea}. For further details on these simulations, including modelling choices for the \HI, see \cite{de_lucia_2007, Obreschkow_2009b, yahya10.1093/mnras/stv695} for S$^3$-SAX, and \cite{de_lucia_2014, 2017MNRAS.465.2236Z, 2024A&A...687A..68D, mayor_2026} for GAEA.

We present the \HI galaxy number density and bias with different constant flux density cuts (represented by the various colours) for the S$^3$-SAX (solid), GAEA (dashed), and TNG100 (dash-dot) \HI simulations in Figure~\ref{fig:dndz}. For comparison, we have also added the \HI bias calculated from the ALFALFA survey at $z\approx 0.022$ \citep{2019MNRAS.486.5124O}. {\edit{We adopt this value as a simple single point of comparison, but note that further \HI galaxy clustering measurements from observations \citep[e.g.][]{2011MNRAS.412L..50P, 2013ApJ...776...43P, 2017ApJ...846...61G, 2018MNRAS.479.1627P} and simulations \citep[e.g.][]{2017MNRAS.465.2236Z, fontanot_2025} are available in the literature.}}

\begin{figure}
    \centering
    \includegraphics[width=\columnwidth]{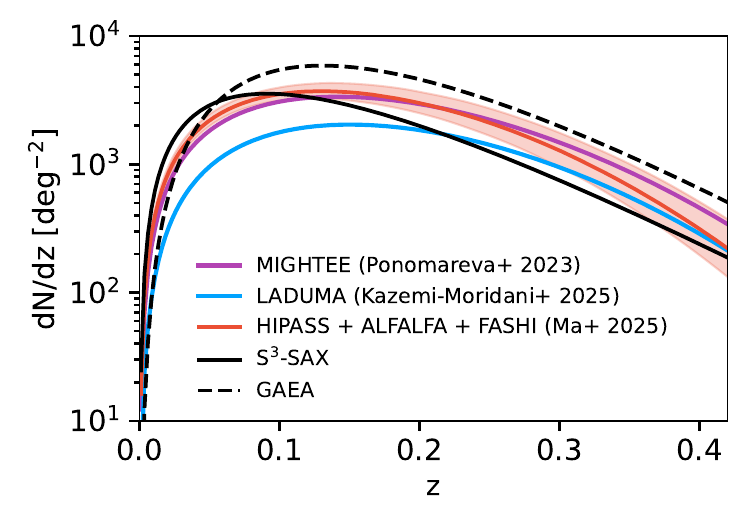}
    \caption{{\edit Comparison of predicted number densities for galaxies with $S_{\rm rms} > 100\,\mu$Jy, assuming a fixed reference line width of 200~km\,s$^{-1}$. The coloured lines show predictions based on \HI mass function (Schechter function) fits to several low-redshift observational datasets \citep{2023MNRAS.522.5308P, KazemiMoridani2025, 2025A&A...695A.241M}, while the black lines show the predictions from S$^3$-SAX and GAEA, as described above. The quoted uncertainties on the Schechter function parameters from \citet{2025A&A...695A.241M} have been used to indicate the uncertainty (95\% CL) in this curve.}}
    \label{fig:dndz_models}
\end{figure}

\begin{figure*}
    \centering
    \includegraphics[width=0.49\linewidth]{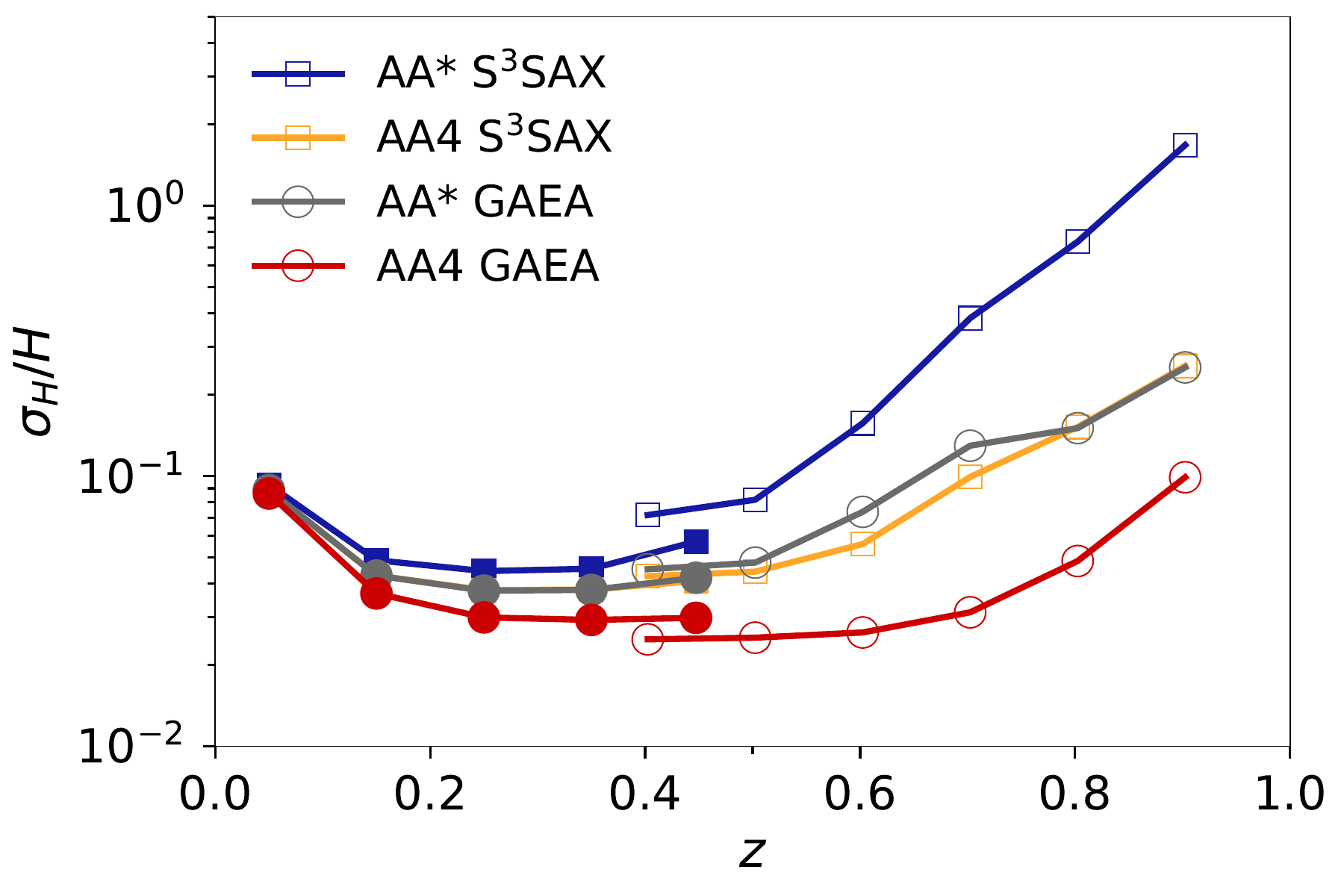}
     \includegraphics[width=0.49\linewidth]{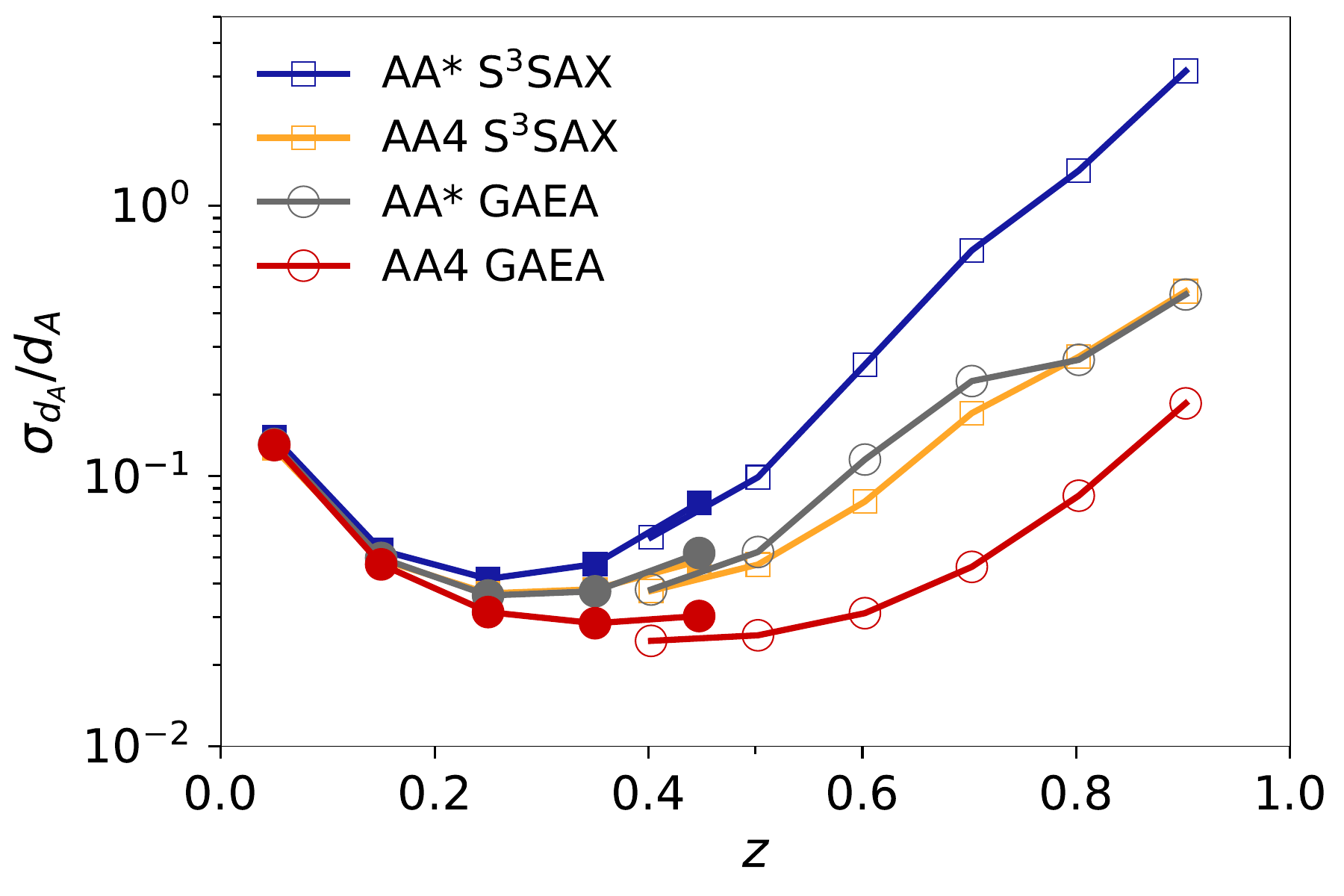}
    \caption{Forecast fractional errors on the expansion rate, $H(z)$, (left) and angular diameter distance, $d_A(z)$, (right) for the AA* and AA4 survey parameters, based on predicted number densities from S$^3$-SAX and GAEA {\edit shown in Tables~\ref{tbl:params_s3sax} and \ref{tbl:params_gaea}}. The filled markers are for Band 2 and the unfilled ones are for Band 1.}
    \label{fig:hz_da}
\end{figure*}

At the $0~\mu$Jy threshold, S$^3$-SAX predicts the highest number density of \HI galaxies across all redshifts, as apparent in Figure~\ref{fig:dndz}. Interestingly, the TNG100 results are comparable to S$^3$-SAX, even though the latter is based on a Semi-Analytical Model (SAM) and the former is a hydrodynamical simulation. In contrast, GAEA differs from the S$^3$-SAX results by a factor of 2--3 at low redshifts, and almost an order of magnitude at higher redshift, despite it being based on the same dark matter-only simulation, Millennium \citep{2005Natur.435..629S}, and using the direct successor of the original SAM \citep{de_lucia_2007} used to generate the S$^3$-SAX catalogue. We also note that the free parameters from S$^3$-SAX used in the fitting function in Equation \ref{eq:dNdz&bias} result in an unwanted behaviour at low $z$ in which the number density at $S_{\rm rms}>100\, \mu$Jy approaches the same values as $S_{\rm rms}=0\, \mu$Jy.

At higher flux density rms threshold values of $S_{\rm rms}>50\,\mu$Jy, the number densities are generally highest from TNG100 and lowest from S$^3$-SAX, with results from GAEA being in between the two simulations. All three simulations have comparable peaks within roughly a factor of two at $z\sim0.1$, but at higher redshifts, the discrepancy can be as large as several orders of magnitude.

In contrast to the number density, the galaxy bias shown in Figure~\ref{fig:bz} has a wide range of values especially at higher flux density rms thresholds. In fact, the difference at $S_{\rm rms}>100\,\mu$Jy, can be up to a factor of 6 for the higher redshifts. It is worth noting that S$^3$-SAX and GAEA show comparable values of the bias, while TNG100 consistently shows bias values of $\sim 0.8$ at $z \lesssim 0.4$. This outcome may be caused by the smaller volume of $\sim(100$ Mpc$)^3$ used by TNG100, making it harder to get a robust estimate of the bias. Hence, we have chosen to only run the cosmological forecasts using S$^3$-SAX and GAEA in the next section. For comparison, the inferred bias from the ALFALFA survey is larger than the bias from all simulations, but has a significantly higher flux density threshold than we considered, at $S_{\rm rms}=2.4$~mJy.

{\edit Fig.~\ref{fig:dndz_models} compares the S$^3$-SAX and GAEA number density predictions with predictions based on observational \HI mass function model fits from MIGHTEE \citep{2023MNRAS.522.5308P}, LADUMA \citep{KazemiMoridani2025}, and a combined analysis of HIPASS, ALFALFA, and FASHI \citep{2025A&A...695A.241M}. These studies are based on galaxy sample sizes of 203, 82, and $\sim$40,000 respectively. In the latter case, the combination of multiple surveys results in a sample with a highly variable depth and angular resolution across the survey area. For the purposes of comparison, we have taken $S_{\rm rms} > 100~\mu$Jy and a fixed line width of 200~km\,s$^{-1}$. The observational fits are based on galaxy samples in the local Universe, i.e. $z \lesssim 0.1$, and so the redshift dependence is an extrapolation without assuming evolution of the mass function. We took the best-fit Schechter function in each case, assuming a standard (single power-law) Schechter function for the \citet{2025A&A...695A.241M} result instead of their double power-law fit.}

{\edit It can be seen from Fig.~\ref{fig:dndz_models} that the spread in the observational model fit predictions is similar to that of the simulation-based predictions. The GAEA prediction has the most similar shape to the observational fits, but a higher number density above $z \simeq 0.05$ and a lower one below this. The S$^3$-SAX prediction peaks at a lower redshift (around $z \approx 0.05$, compared to $z \approx 0.1$ for the other curves), but is similar in amplitude around $z \approx 0.1$.}

{\edit Overall, the number count predictions from the observational and simulation-based studies are consistent to within a factor of a few, and show a similar trend with redshift. We reiterate that the former are based on an extrapolation from $z \simeq 0$ however. Stacking detections at higher redshift \citep[e.g.][]{2022ApJ...940L..10B, 2024MNRAS.529.4192S, 2025MNRAS.544.1710P}, as well as the growing numbers of direct \HI galaxy detections at $z \gg 0.1$ \citep[e.g.][]{2024ApJ...966L..36X, 2025MNRAS.544..193J, 2026ApJ...998..248B} and \HI intensity mapping constraints \citep{2023arXiv230111943P} are starting to place additional constraints on the redshift evolution of the \HI mass function. Stacking studies \citep{2022ApJ...940L..10B, 2025MNRAS.544.1710P} suggest a significant evolution in the \HI mass function out to $z \sim 0.5$ for example, particularly for massive galaxies. We have not attempted to construct a redshift-dependent \HI mass function model fit to the observations here however.}

\section{Fisher forecasts for cosmology}
\label{sec:cosmo_fisher}
In this section, we provide a short description of a mathematical formalism for performing Fisher matrix forecasts for galaxy surveys, and discuss the results of the forecast performed using inputs from the S$^3$-SAX and GAEA simulations.

First, we briefly summarise the Fisher matrix formalism for predicting the (Gaussian) uncertainties on cosmological parameters derived from measurements of the galaxy power spectrum. Further detailed are provided in \citet{Bull_2015} and references therein.

The total power spectrum at redshift $z$ and wavevector $\mathbf{k}=(\mathbf{k}_\perp, k_\parallel)$, $P_{\rm tot}(z, \mathbf{k})$, is defined as 
\begin{equation}
    P_{\rm tot}(z, \mathbf{k}) = (b(z)+f \mu^2)^2 \exp (-k^2 \mu^2 \sigma^2_{\rm NL}) D^2(z) P(z=0, \mathbf{k}),
\end{equation}
where $f$ is the dimensionless linear growth rate, $\mu \equiv k_\parallel/k$, $D$ is the linear growth factor, and $\sigma^2_{\rm NL}$ is the non-linear dispersion \citep{Bull_2015}.
From this, the Fisher matrix of a galaxy redshift survey for parameters indexed by $(i,j)$ can be written as
\begin{equation}
    F_{ij}=\frac{1}{2} \int_{k_{\rm min}}^{k_{\rm max}} \frac{d^3 k}{(2 \pi)^3}[\partial_i \textrm{ln}P_{\rm tot}(\mathbf{k}) \, \partial_j \textrm{ln}P_{\rm tot}(\mathbf{k}) ] V_{\rm eff}(\mathbf{k}),
\end{equation}
where $V_{\rm eff}(\mathbf{k})$ is the effective survey volume
\citep{1994ApJ...426...23F, Tegmark_1998}. For a comoving survey volume, $V_{\rm bin}$,
\begin{equation}
\begin{split}
    V_{\rm bin} &\simeq r^2(\bar{z})\frac{dr}{dz}\Delta z \Omega_{\,{\rm survey}},\\
    V_{\rm eff}(\mathbf{k}) &= V_{\rm bin}\left[\frac{\bar{n}(z)P_{\rm tot}(\mathbf{k})}{1+\bar{n}(z)P_{\rm tot}(\mathbf{k})}\right]^2.
\end{split}
\end{equation}
Here, $r$ is the comoving distance, $\bar{z}$ is the mean redshift, $\Omega_{\,{\rm survey}}$ is the survey area, and $n(z)$ is the comoving galaxy number density as a function of redshift. The Fisher matrix is evaluated at a particular `fiducial' cosmological model, which we have taken to be a Planck 2018 flat $\Lambda$CDM cosmology \citep{aghanim2020planck}.

The Fisher matrix can then be inverted to provide a prediction of the covariance matrix for the chosen parameters, i.e. $C_{ij} = (F^{-1})_{ij}$. For these forecasts, we work with parametrised redshift-dependent cosmological functions instead of cosmological parameters, as these are less model-dependent. In each redshift bin, we allow the expansion rate $H(z)$, angular diameter distance $d_{A}(z)$, linear growth rate $f\sigma_8(z)$, and bias $b \sigma_8(z)$ to be free parameters. We also include $\sigma_{\rm NL}$, which is marginalised over as a nuisance parameter to account for uncertainties in the modelling of non-linear scales. We do not include other modelling uncertainties, such as higher-order bias terms. Note that the full anisotropic shape of the \HI galaxy power spectrum is used to derive the constraints, and not just individual features such as the BAO or RSDs.

We use the \texttt{RadioFisher}\footnote{\url{https://github.com/philbull/RadioFisher}}
code \citep{Bull_2015} to calculate the Fisher matrix, using galaxy number count and bias values for 10,000 hours of observation using the SKA-MID AA* and AA4 configurations, and a survey area of 5,000 deg$^2$. We derive these values from the S$^3$-SAX and GAEA predictions presented above to give some sense of how the forecasts depend on the uncertain survey performance estimates.

\begin{figure*}
    \centering
    \includegraphics[width=\linewidth]{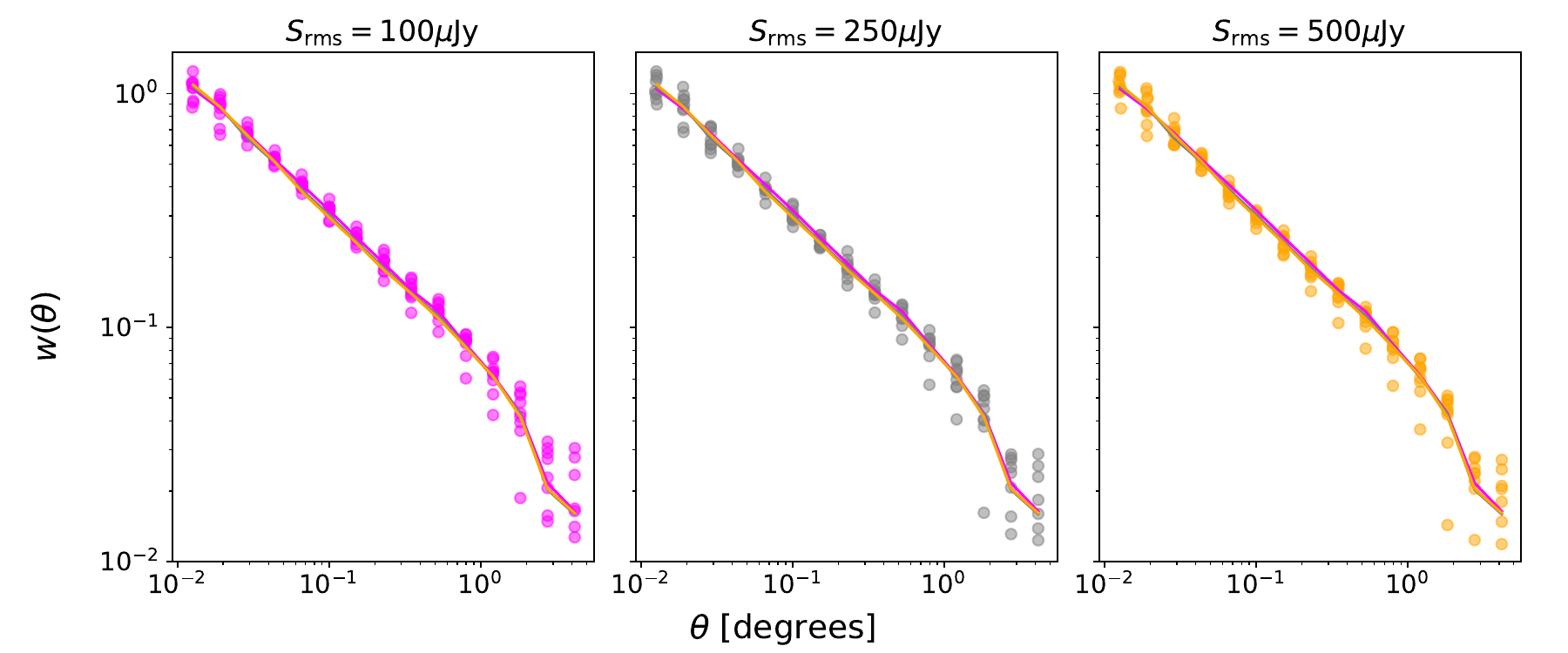}
    \caption{The angular correlation function (ACF) and variance in the 9 fields with various flux density cuts. For each angular bin, there are nine points, one for each field. The mean ACFs for the four flux density cuts are plotted in the corresponding colours in all of the panels to aid comparison; they are almost overlapping.}
    \label{fig:all_acf}
  
\end{figure*}
    
The fractional error on $H(z)$ is shown in the left panel of Figure~\ref{fig:hz_da}. The filled markers are for Band~2 and the unfilled ones are for Band~1. At $z<0.1$, the constraints are degraded, suggesting that they are limited by sample variance, hence a larger survey area would be needed. Overall, the AA4 configuration with the GAEA simulation consistently gives constraints that are better than 10$\%$ in both frequency bands, dropping as low as 2--3\% at $z \approx 0.4$. The other configurations/simulations also give constraints that are better than 10$\%$, but only for $z \leq 0.5$; beyond that redshift, the fractional uncertainty can be up to 11 times larger, as in the case of the AA* S$^3$-SAX prediction at $z\sim0.9$.

With the AA4 configuration, the differences in fractional errors on the cosmological parameters between the two simulations are within 3--4\% in Band 1, while they are within 5--10\% for the AA* configuration in the same band. In other words, the performance differs by a factor of several depending on the simulation used. The fractional error on $d_A(z)$ is generally similar to that of $H(z)$, as shown in the right panel, with only minor differences in the values. 

It should be noted that, even with the relatively optimistic assumptions that the full shape of the power spectrum can be modelled, and that 10,000 hours of survey time can be used, the forecast constraints from SKA-MID surveys are significantly poorer than comparable optical galaxy surveys. The most directly comparable is the DESI Bright Galaxy Survey \citep[BGS;][]{2023AJ....165..253H}, which operates at similar redshifts, but will ultimately cover 14,000 deg$^2$ with $\sim 10^3$ galaxies per square degree. Its very wide survey area goes some way to mitigating the small cosmic volumes available at low redshift.

Ultimately, greater survey volumes (i.e. more available Fourier modes) are needed to beat down sample variance and improve the cosmological constraints, and this can only be achieved efficiently by going to higher redshifts, e.g. $z \sim 1 - 2$, as targeted by emission line galaxy surveys at optical wavelengths. The predicted \HI galaxy number densities for wide SKA-MID surveys (flux density rms around 100~$\mu$Jy) are too low at these redshifts however. A much larger array -- an ``SKA2'' class instrument \citep{yahya10.1093/mnras/stv695, Bull:2015lja} -- would be needed to achieve a flux density rms of $\lesssim 10~\mu$Jy or below, resulting in competitive cosmological constraints, e.g. at the sub-percent level on $H(z)$ and $d_A(z)$.

\section{Angular Correlation Function and Halo Occupation Distribution}
\label{sec:acf_hod}

In this section, we make use of the TNG300 dataset at snapshot $z=0$ to construct nine different realisations of mock \HI galaxy redshift surveys over a square field of view of side length 20$^\circ$, as shown in Figure~\ref{fig:hi_galaxies}. For each realisation, we set the flux density threshold to $S_{\rm rms} = [100, 250, 500]\, \mu$Jy, calculate the angular correlation functions of galaxies above these thresholds, and fit the functions using a halo model to infer the 1- and 2-halo terms of the power spectrum, given by
\begin{equation}
    \begin{split}
        P_{\rm 1-halo} &= \frac{1}{\bar{n}_{\rm gal}^2} \int \frac{\textrm{d}n}{\textrm{d}M_h}\textrm{d}M_h \, \bar{N}_c f_c [\bar{N}_s^2u^2_s(k)+ 2\bar{N}_s u_s(k)]\\
        P_{\rm 2-halo} &= \left( \frac{1}{\bar{n}_{\rm gal}} \int \frac{\textrm{d}n}{\textrm{d}M_h} \textrm{d}M_h \,b(M_h) \bar{N}_c [f_c + \bar{N}_su_s(k)] \right)^2 {P_{\rm lin}(k)}.
    \end{split}
\end{equation}
Here, $\bar{n}_{\rm gal}$ is the mean galaxy number density, ${\textrm{d}n}/{\textrm{d}M_h}$ is the halo mass function, $\bar{N}_c$ is the mean number of central galaxies, $\bar{N}_s$ is the mean number of satellite galaxies, $u_s(k)$ is the normalised satellite galaxy density profile, $f_c$ is the observed fraction of central galaxies, and $P_{\rm lin}(k)$ is the linear matter power spectrum \citep{Nicola_2020, ye-acf10.1093/mnras/staf1651}. The sum of the 1- and 2-halo terms give the full galaxy power spectrum, $P(k)$. The bias can then be calculated from the full power spectrum as
\begin{equation}
\label{eqn:b_gal}
    b_{\rm gal}^2(k) = \frac{P(k)}{P_{\rm lin}(k)}.
\end{equation}

We briefly describe the angular two-point correlation function and its covariance matrix in Section~\ref{sec:acf_cov}, and the Bayesian inference method we have applied in Section~\ref{sec:hod_fitting}. We then discuss the results in Section~\ref{sec:results_acf}.

\begin{figure*}
    \centering
    \includegraphics[width=0.88\linewidth]{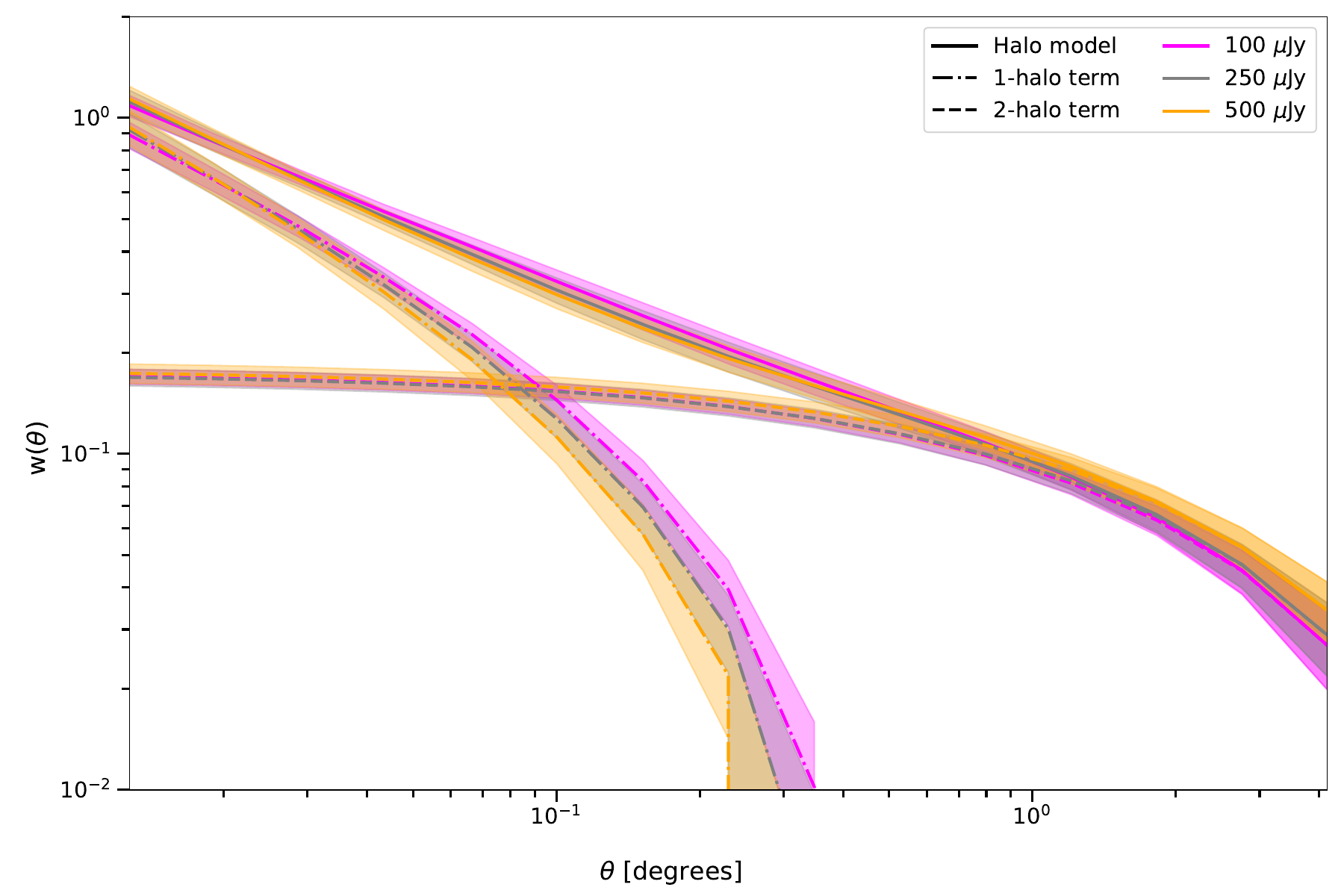}
    \caption{The halo model predictions for the 1-halo term (dash dot line), 2-halo term (dashed line), and the full halo model (solid line) with the various flux density cuts. The curves shown are averages of the posterior means from the nine fields, while the shaded regions show the scatter (standard deviation) over the nine fields. The integral constraint and shot noise are not included here (although they are included when performing the inference in each field).}
    \label{fig:1h_2h}  
\end{figure*}

\subsection{Angular correlation function and its covariance}
\label{sec:acf_cov}

The angular correlation function $w(\theta)$ is defined as
\begin{equation}
    \label{eqn:acf}
    w(\theta) = \langle \delta_{\rm g} (\theta + \theta_{\rm c}) \, \delta_{\rm g}(\theta_{\rm c})\rangle,
\end{equation}
where $\delta_{\rm g}$ is the galaxy overdensity, $\theta$ is the angular separation, and $\theta_{\rm c}$ is a reference angular position. It quantifies the projected clustering of galaxies compared to a random Poisson distribution. If the pair counts are performed in discrete angular bins, the equation is effectively
\begin{equation}
    \label{eqn:acf_bin}
    w(\theta_\alpha) = \frac{\sum_{p,q} \delta_{\rm g}^{(p)} \delta_{\rm g}^{(q)} \Theta^\alpha_{pq}}{\sum_{p,q} \Theta^\alpha_{pq}}.
\end{equation}
Here, $\alpha$ is the angular bin, $p$ and $q$ are the unique pairs of \HI galaxies in the field, and $\Theta^\alpha_{pq}$ is an indicator function where\\
$ \Theta^\alpha_{pq} = 
\begin{cases}
1\hspace{0.5cm} \text{if } \theta \in \alpha\\
0\hspace{0.5cm} \text{if } \theta \notin \alpha.
\end{cases}$\\

To take into account possible limitations in the data, e.g. masked regions and variable depth, statistical estimators such as the Peebles \citep{peebles-hauser1974ApJS...28...19P} or Landy-Szalay estimators \citep{landy-szalay1993ApJ...412...64L} are commonly used to measure $w(\theta)$. Because of its lower variance and bias, we have chosen to use the Landy-Szalay estimator, which is given by
\begin{equation}
    w(\theta) = \frac{N_{dd}-2N_{dr} + N_{rr}}{N_{rr}}.
\end{equation}
For a binned angular separation $\theta$, $N_{dd}$ is the number of galaxy-galaxy pairs, $N_{dr}$ is the number galaxy-random pairs, and $N_{dr}$ is the number of random-random pairs. `Randoms' are catalogues of points placed at random according to a Poisson distribution, but that have the same mask/selection function as the real galaxies. In this work, we use the \textsc{TreeCorr} package \citep{2004MNRAS.352..338J, treecorr2015ascl.soft08007J} to estimate the angular correlation function in 15 angular separation bins that are equally spaced on a logarithmic scale between 0.01$^\circ$ and 2$^\circ$. The random catalogue is generated following a uniform distribution between $-10^\circ$ and $10^\circ$. In each field, the total number of randoms being generated is 10 times the number of the observed galaxies, to help reduce the variance of the estimator.

Next, we use the jackknife method to calculate the covariance matrix of the angular correlation function, $C$, given by
\begin{equation}
    C = \frac{N_{\rm patch}-1}{N_{\rm patch}}\sum_i (w_i -\bar{w})^{\rm T}(w_i -\bar{w}).
\end{equation}
Here, $N_{\rm patch}$ is the number of patches that we have sub-divided each survey area into, $w_i$ is the estimated correlation function excluding the $i$'th patch, and $\bar{w}$ is the mean correlation function across all patches. In our work, we have set $N_{\rm patch}=500$ to reduce the variance on the estimated covariance matrix.

We present the angular correlation function in each of the nine fields with flux density cuts of 100 (left panel), 250 (middle panel), and 500~$\mu$Jy (right panel) in Figure~\ref{fig:all_acf}. {\edit At separations below $1^\circ$, the values of the angular correlation functions across the nine fields are consistently within a factor of 3, and there is very little evolution with flux density cut.}

For reference, in Appendix~\ref{app:correlation} we also present correlation matrices for the angular correlation function bins in each of the nine fields for galaxies brighter than 100 $\mu$Jy. These show that our jackknife covariance estimation procedure is appropriate, and that the logarithmic binning in separation produces almost-independent correlation function bins, with only mild correlations between neighbouring bins at separations of tens of arcminutes. 

\begin{figure*}
    \centering
    \includegraphics[width=\linewidth]{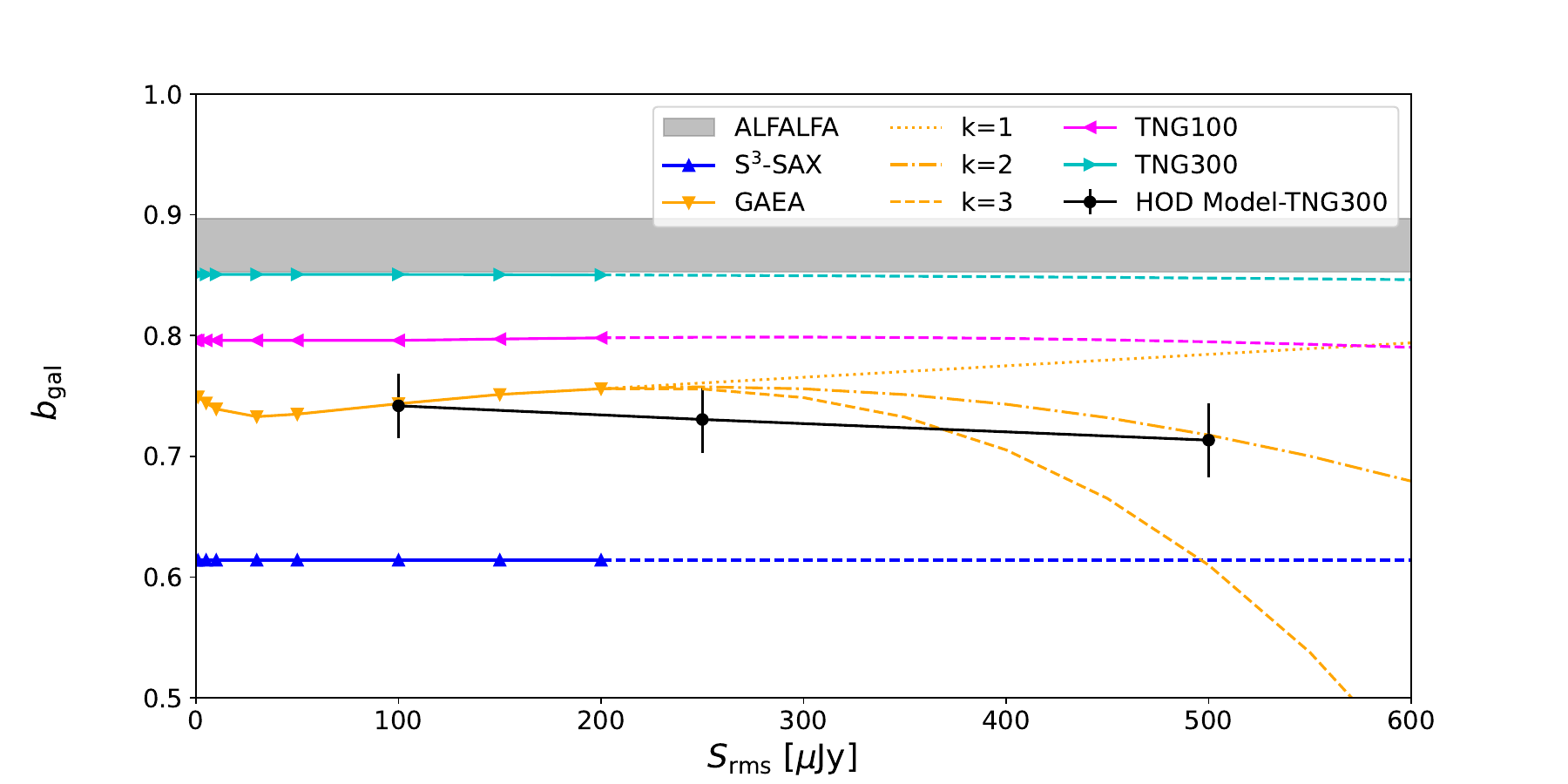}
    \caption{The mean and uncertainty of the galaxy bias from the halo model (circle) and \HI simulations (triangles as labelled) as a function of flux density threshold at $z=0.05$. The mean and uncertainty are calculated from the mean and standard deviation of the fitted bias of the nine fields. The shaded region shows the observed bias with $S_{\rm rms}=2.4$ mJy from the ALFALFA survey \citep{2019MNRAS.486.5124O}. The dotted, dashed-dot, and dashed lines are extrapolated values at higher $S_{\rm rms}$.}
    \label{fig:bias}  
\end{figure*}

\subsection{Bayesian parameter inference for the HOD model}
\label{sec:hod_fitting} 

To fit the halo occupation distribution to the angular correlation function, we use the same method as \cite{ye-acf10.1093/mnras/staf1651}. Using the halo occupation distribution model from \textsc{CCL} \citep{Chisari_2019}, we vary four parameters, $\mathbf{p}$ = [log$_{10} M_{\rm min}$, log$_{10} M_{1}$, $f_c$, $\beta_g$]. These represent the logarithm of the minimum mass halo that can host a central galaxy; the logarithm of the {\edit minimum halo mass for which  at least} one satellite galaxy is hosted by a halo; the observed fraction of central galaxies; and a quantity that relates the spatial distribution of satellite galaxies to their host haloes. The values of the model angular correlation function $w^{\rm m}(\theta, \mathbf{p})$ and galaxy number density $n_{\rm g}^{\rm m}(\mathbf{p})$ calculated from a given set of parameters are then used to evaluate the likelihood function in a Bayesian Markov Chain Monte Carlo (MCMC) setting using the \textsc{emcee} package \citep{emcee2013ascl.soft03002F}. The likelihood function is given by
\begin{align}
    \mathrm{log} \mathcal{L} & = - \frac{1}{2} \left( \sum_{i,j} [w^{\rm d}(\theta_i) - w^{\rm m}(\theta_i, \mathbf{p}) ]\, C_{ij}^{-1}\, [w^{\rm d}(\theta_j) - w^{\rm m}(\theta_j, \mathbf{p}) ] \right . \nonumber\\
    &~~~~~~~~~~~~~~~+ \left . \frac{[\mathrm{log} \,n_{\rm gal}^{\rm d} - \mathrm{log} \,n_{\rm gal}^{\rm theor}(\mathbf{p}) ]^2}{\sigma^2_{\mathrm{log} \, n_{\rm gal} }} \right), \label{eq:logL}
\end{align}
where the superscripts d and m correspond to the values for data and model respectively, $n_{\rm gal}^{\rm theor}$ is the theoretical number density from the halo model, and $\sigma_{\mathrm{log} \, n_{\rm gal} }$ is the standard deviation of ${\mathrm{log} \, n_{\rm gal} }$, chosen to be 10\% of ${\mathrm{log} \, n_{\rm gal}^{\rm d} }$. This second term is an ad hoc regularisation term intended to penalise models that predict strongly incorrect number counts. Here, $w^{\rm m}(\theta_i, \mathbf{p})$ is related to the theoretical value from the halo model, $w^{\rm theor}(\theta_i, \mathbf{p})$, following
\begin{equation}
    w^{\rm m}(\theta_i, \mathbf{p}) = w^{\rm theor}(\theta_i, \mathbf{p}) - \frac{\sum_i N_{rr} w^{\rm d}(\theta_i)}{N_{rr}} + \frac{1}{N_{\rm gal}},
\end{equation}
where $N_{\rm gal}$ is the total number of galaxies in the comoving volume. The second term here is the integral constraint \citep[IC;][]{1977ApJ...217..385G}, added to account for finite volume effects, and the final term is the shot noise.

We use the same priors described in \cite{ye-acf10.1093/mnras/staf1651} for our four parameters, and run the MCMC with 32 walkers and 10,000 iterations. We discard the first 20$\%$ of each chain as burn-in, and visually inspect the rest to ensure they have reached acceptable convergence.


\begin{table}
\begin{tabular}{|c|c|c|c|c|}
\hline
$S_{\rm rms}$ & log$_{10} M_{\rm min}$ & log$_{10} M_{1}$ & $f_c$            & $\beta_g$        \\ \hline
100~$\mu$Jy                       & 10.21 $\pm$ 0.12        & 12.56 $\pm$ 0.05 & 0.47 $\pm$ 0.10 & 0.205 $\pm$ 0.043 \\ \hline
250~$\mu$Jy                        & 10.16 $\pm$ 0.13        & 12.66 $\pm$ 0.05 & 0.39 $\pm$ 0.10 & 0.137 $\pm$ 0.030 \\ \hline
500~$\mu$Jy                        & 10.02 $\pm$ 0.17        & 12.82 $\pm$ 0.06 & 0.14 $\pm$ 0.06 & 0.093$\pm$ 0.022  \\ \hline
\end{tabular}
\caption{{\edit \label{tbl:hod_params} The weighted mean and standard deviation of the parameters $\log_{10} M_{\rm min}$, $\log_{10} M_{1}$, $f_c$, and $\beta_g$ over eight of the nine sky realisations with three different flux density thresholds.}}
\end{table}


\begin{figure*}
    \centering
    \includegraphics[width=0.98\linewidth]{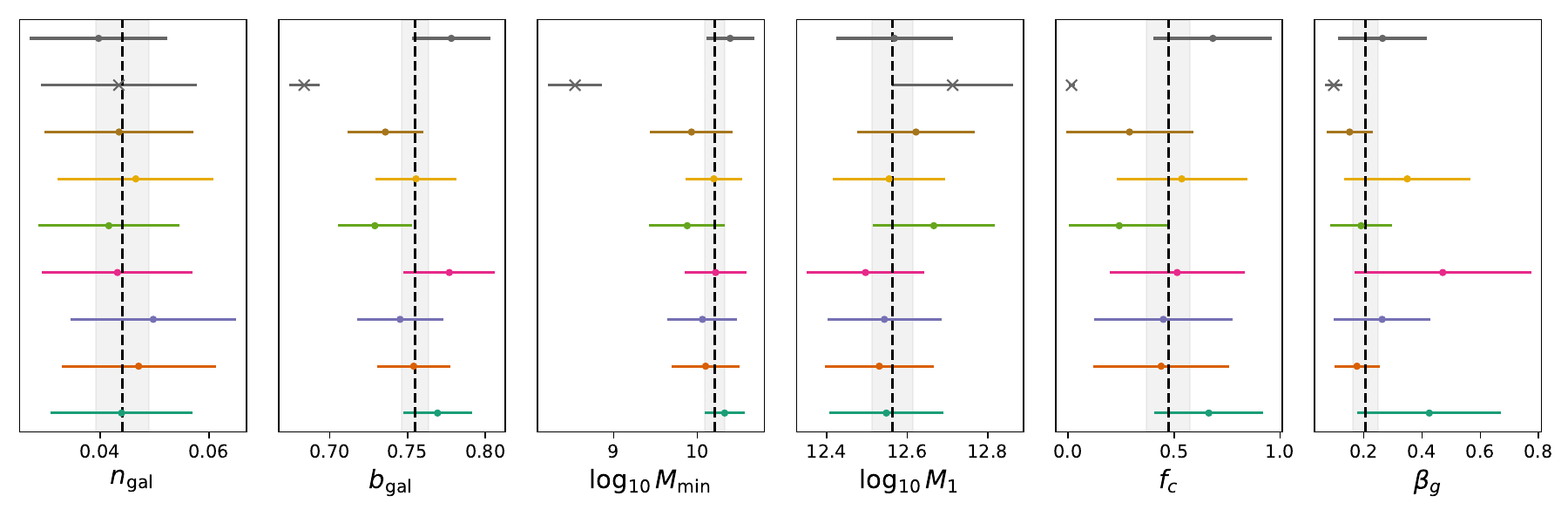}
    \caption{{\edit The marginalised 1D posterior values of the parameters for all nine realisations with $S_{\rm rms}=100\mu$Jy. The vertical dashed line and grey band denote the weighted mean and standard deviation of eight of the nine points (excluding the outlier point, second from the top, marked with an `$\times$'). The units of $n_{\rm gal}$ are $10^{-3}\,{\rm Mpc}^{-3}$.}}
    \label{fig:marginal_values}  
\end{figure*}

\subsection{Parameter inference results}
\label{sec:results_acf}

Figure~\ref{fig:1h_2h} shows the mean halo model predictions for the 1-halo term (dash-dot line), 2-halo term (dashed line), and the full halo model (solid line) as well as their uncertainty (shaded region), for the various flux density cuts. To clarify, the mean and uncertainty have been calculated over the posterior means for each of the nine fields. In general, there is minimal difference in the angular correlation function of the 2-halo term for all three flux density thresholds, except at $\theta \geq 0.5^\circ$. In contrast, the angular correlation function of the 1-halo term starts to show differences as a function of flux density threshold at $\theta \sim 0.03^\circ$, going up to a factor of 5 difference at $\theta \sim 0.2^\circ$.


Figure~\ref{fig:bias} shows the galaxy bias, $b_{\rm gal}$, computed from the posterior mean HOD model (circle), compared with the values estimated from the \HI simulations (triangles) as a function of flux density threshold. We also include the observed bias at $S_{\rm rms}=2.4$~mJy from the ALFALFA survey, marked as a shaded region. Similar to the halo model, the mean and uncertainty of the galaxy bias are calculated from the mean and standard deviation of the fitted bias in the nine fields. The HOD model bias value has been calculated following Equation \ref{eqn:b_gal}, in which the posterior mean values of the HOD parameters are used to calculate the full power spectrum. The values for the \HI simulations are taken from the fitting functions (points and solid lines), while the dotted, dashed-dot, and dashed lines show extrapolated values of these out to higher $S_{\rm rms}$. While most are very flat with flux density threshold, the GAEA results are not, and so we use polynomial fits with different degrees to extrapolate, to highlight the uncertainty in this extrapolation. 

It can be seen that $b_{\rm gal}$ from the HOD model has a slightly downward trend as the flux density threshold increases, with a change of $\Delta b_{\rm gal}=-0.05$ as the flux density threshold increases by an order of magnitude. This could be caused in part by the lack of redshift evolution of the haloes, as we are only using the data at one snapshot, for smaller sub-volumes of the 300 Mpc/$h$ simulation. The change in mean halo mass as a function of flux density threshold was also studied in Sect.~\ref{sec:TNG300}, where comparatively little evolution was found. 
The $b_{\rm gal}$ values inferred from the HOD model inference are closest to the results from GAEA, although the values for the other simulations are within about 20\% of this.

\begin{figure*}
    \centering
    \includegraphics[width=0.95\linewidth]{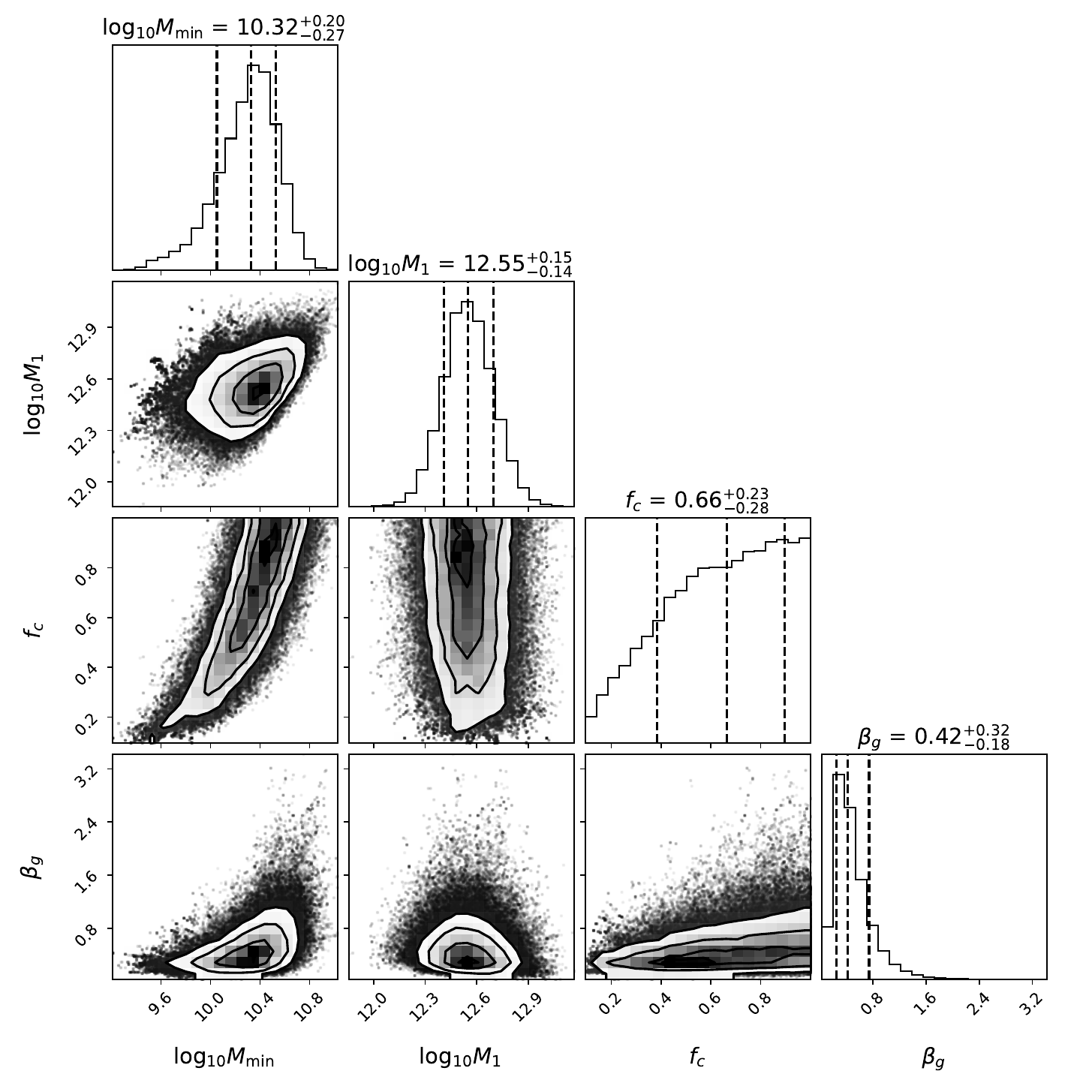}
    \caption{Corner plot of the marginalised posterior distribution of the parameters log$_{10} M_{\rm min}$, log$_{10} M_{1}$, $f_c$, and $\beta_g$ for one sky realisation with $S_{\rm rms}=100\mu$Jy {\edit (the realisation represented by the bottom-most points in Fig.~\ref{fig:marginal_values}).}}
    \label{fig:corner_mcmc}  
\end{figure*}

{\edit We present summary statistics of the four free HOD model parameters with three different flux density thresholds in Table~\ref{tbl:hod_params}. The values are calculated from a weighted average of the posterior mean values for each parameter from eight of the nine sky realisations. One realisation is excluded as the MCMC chain exhibited poor convergence for the $f_c$ parameter. This can be seen as a clear outlier in Fig.~\ref{fig:marginal_values}, which shows the posterior mean values for each realisation (coloured points) for $S_{\rm rms} = 100\,\mu$Jy, along with the weighted mean and standard deviation (vertical dashed line and shaded area respectively). This figure also shows values for the galaxy number density and bias. To more clearly show the correlations between parameters, we also include an example corner plot of the marginal posterior distribution for one sky realisation with $S_{\rm rms}=100\,\mu$Jy in Figure~\ref{fig:corner_mcmc}.}

{\edit According to Table~\ref{tbl:hod_params}, the mean values of $\log_{10} M_{\rm min}$, $f_c$, and $\beta_g$ decrease as the flux density threshold increases. If we interpret $f_c$ as a duty cycle parameter, this implies that almost half of central galaxies should be observable for $S_{\rm rms} = 100\,\mu$Jy. This is a departure from current \HI galaxy surveys, which predominantly see gas-rich satellite galaxies. The larger values of $\beta_g$ for smaller $S_{\rm rms}$ imply that a more extended spatial distribution of satellite galaxies is observed within the host dark matter halo. The mass of halos hosting an average of one satellite galaxy per halo, $\log_{10} M_{1}$, decreases as $S_{\rm rms}$ decreases, as expected for a deeper survey. The trend in $\log_{10} M_{\rm min}$ runs counter to this however -- the minimum mass threshold for a halo to contain a central galaxy is actually increasing with increasing survey depth. The correlation between $f_c$ and $\log_{10} M_{\rm min}$ observed in Fig.~\ref{fig:corner_mcmc} may partly explain this; the trend in $\log_{10} M_{\rm min}$ is relatively weak and $f_c$ is poorly constrained, so this trend is possibly just a statistical fluctuation driven by the drift in $f_c$.}

{\edit Despite similar conditions in each realisation, there is significant diversity in the posterior mean values of the HOD parameters. Substantial differences in the posterior means of $f_c$ and $\beta_g$ are observed between realisations for instance. The variation is mostly consistent with the estimated uncertainties however, with no realisations (apart from the discarded one) showing any serious tension with one another. The sub-volumes of the simulation are immediately adjacent to one another, and so some degree of correlation is expected, which may explain the relatively small scatter of the posterior means compared with the size of the errorbars (although for $n_{\rm gal}$ this is more likely to be caused by how the ad hoc regularisation term was included in Eq.~\ref{eq:logL} instead). The main conclusion that we draw from these results is that a medium-deep \HI galaxy survey with SKA-MID should be sufficient to reliably measure the clustering properties of these galaxies without major sample variance effects, although the uncertainties on the HOD parameters will remain substantial.}


Given the relative insensitivity of the clustering results to flux density threshold, we suggest that the model parameters collated in Table~\ref{tbl:hod_params} should be suitable for use as fiducial parameter values for other theoretical HOD model predictions of low-redshift \HI galaxy clustering observables.

\section{Conclusions}
\label{sec:conclusion}

With radio telescope arrays now reaching the sensitivity required to perform large cosmological surveys of galaxies -- numbering in the hundreds of thousands or more -- it is useful to be able to predict their performance to see how well they should be able to constrain cosmological parameters. The main quantities needed for such forecasts are the number density of galaxies as a function of redshift, and the galaxy bias with respect to the underlying dark matter field. The former sets the shot noise level, while the latter sets the overall amplitude of the galaxy power spectrum.

Radio galaxies are easiest to detect through their continuum emission, but this lacks spectral features that would be suitable for estimating redshifts. The most promising {\it spectroscopic} galaxy redshift tracer at radio wavelengths is the 21cm line from neutral hydrogen (\HI) embedded within the galaxies. Indeed, the prospect of observing such galaxies on a massive scale, all the way out to $z \sim 10$, was one of the early drivers of what has now become the Square Kilometre Array Observatory \citep{1991ASPC...19..428W}. It is very faint however, and to date only a few galaxies have been observed through this line beyond the local Universe. As a result, our observational knowledge of the number counts and bias of\HI galaxies is quite poor.

Simulations can be used to plug this knowledge gap, allowing predictions to be made of the numbers of faint galaxies that will become detectable only with forthcoming arrays. In this paper, we have compared the \HI galaxy number count and bias as a function of redshift and sensitivity threshold from three simulations that use different methods to model galaxies: S$^3$-SAX, GAEA, and IllustrisTNG. We found a large spread in their predictions. At low redshift, the number densities disagree with one another by a factor of a few, but this grows to more than an order of magnitude above $z \gtrsim 0.5$ for higher flux density thresholds ($\gtrsim 10\,\mu$Jy), and above $z \gtrsim 1$ for lower flux density thresholds ($\lesssim 10\,\mu$Jy). The bias is also quite uncertain, at the tens of percent level at $z \simeq 0$, rising to about 100\% above $z \gtrsim 1$.

For near-term surveys with arrays such as SKA-MID and its precursors, the theoretical uncertainty in the bias is broadly tolerable, and will affect predictions only at around the $\sim 10\%$ level. For the number counts, however, the large disagreements are potentially more significant.

Due to the poor bias estimate from IllustrisTNG, we only used values from S$^3$-SAX and GAEA to forecast the cosmological performance of a proposed SKA-MID cosmological survey using the Fisher Matrix method. We found that the SKA-MID AA4 configuration using the GAEA simulation's predicted numbers gives constraints on $H(z)$ and $d_A(z)$ that are better than 10$\%$, improving to 2--3$\%$ at $z \approx 0.4$. With the other configurations/simulations, the constraints that are also better than 10$\%$ for $z < 0.5$, but beyond that can degrade significantly, reaching uncertainties that are up to 11 times larger. Comparing the two SKA configurations, with AA4 the differences between the two simulations are within 3--4 $\%$ in Band 1, while the differences are within 5--10 $\%$ for AA* in the same band.

In summary, our results show that the cosmology forecasts at low redshift are mildly dependent on the simulation used to derive the number counts and bias. At higher redshift, $z \gtrsim 0.5$, i.e. SKA-MID Band~1, there are large differences depending on the simulation used. The reason for the insensitivity at low redshift is that sample variance (due to the limited cosmic volume available at low redshift) rapidly becomes the dominant source of uncertainty on the measured two-point statistics, while at higher redshift, the different shot noise levels are more consequential.

To model a {\edit `medium-deep' SKA} \HI galaxy survey at low redshift, we also took sub-samples of galaxies from IllustrisTNG-300 and subjected them to a realistic galaxy clustering analysis, in which the angular correlation function in a single redshift bin is fit using a Halo Occupation Distribution (HOD) model. We performed this analysis separately on nine sub-volumes of the simulations, each spanning a square field of view of 20$^\circ$ on each side, with up to about 20,000 \HI galaxies per sub-volume. Splitting the volume in this way allows us to assess {\edit sample variance effects, which can be important for the smaller survey volumes available at low redshift.}

We found that the measured angular correlation functions had some scatter between sub-volumes, but their average was remarkably consistent over different flux density thresholds. The 1-halo term had the largest spread with respect to flux density threshold, with a factor of 5 difference at an angular separation of $\theta \sim 0.2^\circ$. There is minimal difference in the 2-halo term except at $\theta > 0.5^\circ$ however.
One explanation for this is found in the observation that the mean host halo mass varies quite slowly as a function of flux density. Since the distribution of halo masses is what governs the large-scale clustering behaviour for a galaxy population within the halo model, this may explain the insensitivity to flux density threshold. We also investigated the galaxy bias, $b_{\rm gal}$, finding that it decreased only slightly as a function of flux density threshold, with $\Delta b_{\rm gal} = -6\%$ as the flux density threshold is increased by an order of magnitude. We also compared the HOD-derived bias with the simulation and fitting function predictions, finding agreement to within about 20\% across all of the simulations.

Finally, we explored the inferred means and standard deviations of the HOD parameters, which are $\log_{10} M_{\rm min}$, $\log_{10} M_{1}$, $f_c$, and $\beta_g$. {\edit We observed that most parameters have a downward trend with respect to the flux density threshold, as expected for deeper surveys that include progressively more galaxies. The exception was $\log_{10} M_{1}$, which has a mild upward trend with low statistical significance.}

In conclusion, we have investigated how well different \HI-containing simulations agree with one other, showing that there is substantial disagreement between them in terms of their predictions for \HI galaxy observables. This does not greatly affect cosmological forecasts at the low redshifts relevant to Band~2 of SKA-MID, but becomes much more important at the higher redshifts covered by Band~1. Moreover, the fitting functions used to predict the number density and bias of \HI galaxies may not have an accurate fit at low redshifts, as seems to be the case for S$^3$-SAX for example. As new \HI galaxy survey data arrive, we will have ample opportunity to test and refine these simulation models, but in the meantime, predictions made from such simulations should be assigned a significant theoretical uncertainty.

\section*{Acknowledgments}

We are grateful to J.~Healy and A.~Nicola for useful discussions, {\edit and to J.~Mayor for providing number count and bias results based on the GAEA simulation}. This result is part of a project that has received funding from the European Research Council (ERC) under the European Union's Horizon 2020 research and innovation programme (Grant agreement No. 948764).

\section*{Data Availability}

The simulations that our results are based on are publicly available via their respective references/data repositories. Our analysis and fitting code is available from 
\url{https://github.com/isabelleye1/ACF-halomodel-mcmc/tree/hi_gal}.

\balance



\bibliographystyle{mnras}
\bibliography{hi_galaxies_tng} 




\appendix

\section{Correlation matrices for IllustrisTNG sub-fields}
\label{app:correlation}

\begin{figure*}
    \centering
    \includegraphics[width=0.9\linewidth]{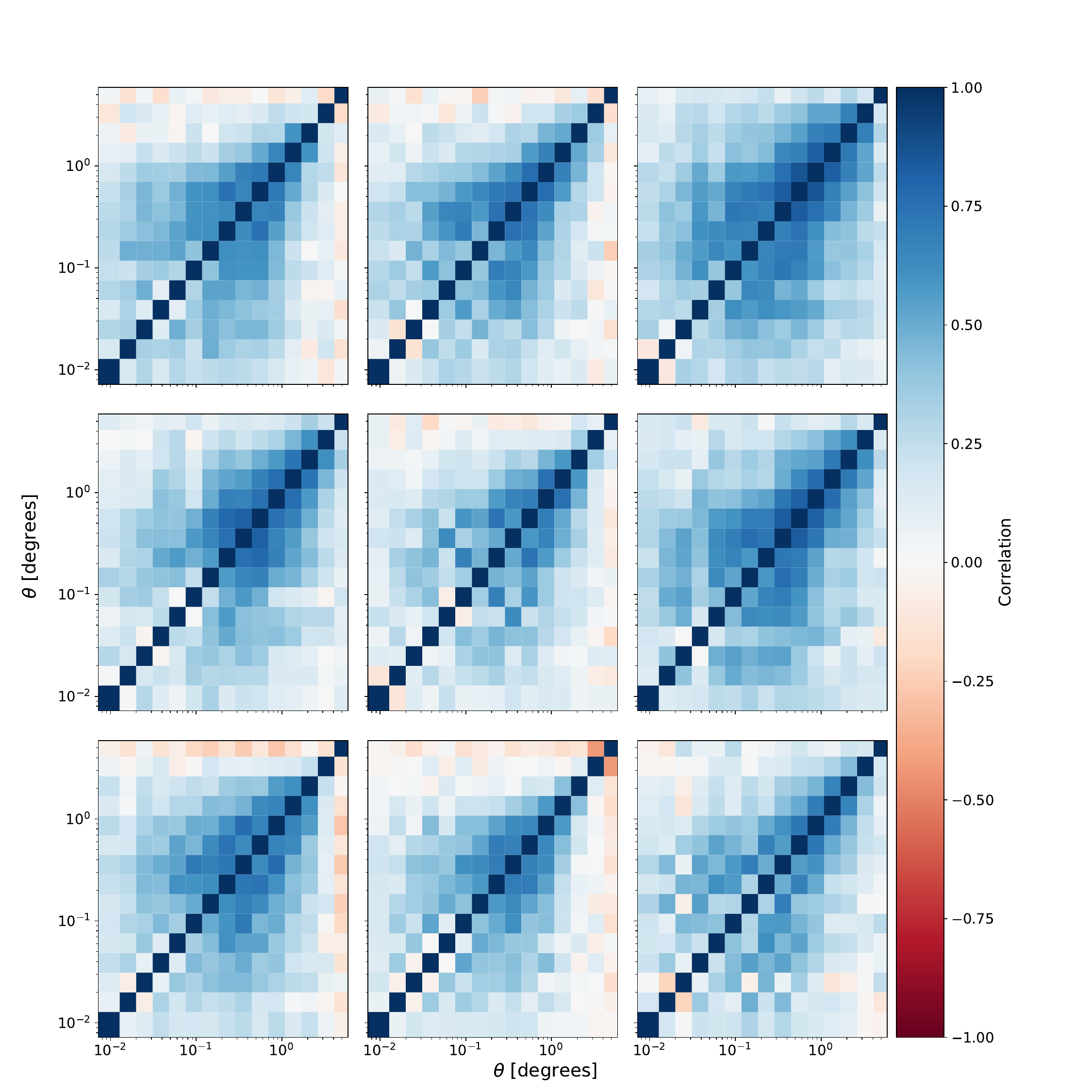}
    \caption{The measured correlation matrices of \HI galaxies brighter than 100 $\mu$Jy in the 9 fields.}
    \label{fig:corr}  
\end{figure*}

In Figure~\ref{fig:corr} we plot correlation matrices for the nine sub-fields of the IllustrisTNG simulation that we analysed in Section~\ref{sec:acf_cov}. These are derived by taking the jackknife estimates of the covariance matrices, and scaling each element by the square root of the product of the parameter variances, i.e. the correlation matrix $\rho_{ij}$ for covariance matrix $C_{ij}$ is defined as
\begin{equation}
\rho_{ij} = \frac{C_{ij}}{\sqrt{C_{ii} \, C_{jj}}}.
\end{equation}
The correlation matrix elements must have values between $+1$ (perfectly correlated), $0$ (uncorrelated), and $-1$ (perfectly anti-correlated).

The general correlation structure of the matrices shown in Figure~\ref{fig:corr} is similar between fields, with relatively little correlation between angular separation bins at the smallest separations (of order arcminutes), and a moderate correlation between bins and their $2-4$ nearest neighbours at intermediate separations of tens of arcminutes. This partly validates the jackknife covariance estimation approach for the angular correlation function for this size of field. It is also notable that there remains noticeable variation however, and so some care would need be taken in accounting for the variance on covariance estimates in a real cosmological \HI galaxy survey.



\bsp	
\label{lastpage}
\end{document}